\documentclass[twocolumn, secnumarabic, amssymb, nobibnotes, pre, superscriptaddress]{revtex4}

\usepackage{graphicx}
\usepackage{amsmath}
\usepackage{float}
\usepackage{color}
\usepackage[english]{babel}
\usepackage{array}
\usepackage[T1]{fontenc}
\usepackage[usenames,dvipsnames]{xcolor}
\usepackage{setspace}
\usepackage{bm}
\usepackage{array} 
\usepackage{makecell} 
\usepackage{lipsum}

\definecolor{darkblue}{rgb}{0,0,0.6}
\definecolor{darkred}{rgb}{0.6,0,0}
\usepackage[colorlinks=true,urlcolor=darkblue,citecolor=darkblue, linkcolor=darkred,hyperfootnotes=false]{hyperref}

\usepackage{tikz}
\usetikzlibrary{arrows.meta}
\usetikzlibrary{decorations.pathreplacing}
\usetikzlibrary{decorations.pathmorphing}
\usetikzlibrary{decorations.markings}
\usetikzlibrary{math}
\usetikzlibrary{tikzmark}
\usepackage{stmaryrd}
\usepackage{stackengine}

\definecolor{color1}{rgb}{0.122,0.467,0.706}
\definecolor{ColorSP}{rgb}{0.925, 0.110, 0.141}
\definecolor{ColorSM}{rgb}{0.259, 0.396, 0.686}
\definecolor{colorFerro}{rgb}{0.957,0.843,0.890}
\definecolor{colorAntiFerro}{rgb}{1.0,0.965,0.835}
\definecolor{colorMixed}{rgb}{1.0,0.902,0.835}
\definecolor{colorPreisach}{rgb}{0.949,0.949,0.949}
\definecolor{color1}{rgb}{0.122,0.467,0.706}
\definecolor{color2}{rgb}{0.839,0.153,0.157}
\definecolor{colorFrozen}{rgb}{0.0,0.75,1.0}
\definecolor{colorSWAP}{rgb}{0.58,0.81,0.58}
\definecolor{newGreenColor}{rgb}{0.24,0.70,0.44}
\definecolor{colorOpen}{rgb}{0.588,0.201,0.541}
\definecolor{colorClosed}{rgb}{0.0,0.039,0.572}

\definecolor{colorCO_new}{rgb}{0.337243401759531,1.0,0.6304985337243401}
\definecolor{colorWW_new}{rgb}{0.9132231404958671,0.0033670033670029076,0.0}

\colorlet{myred}{red!70!black}
\colorlet{xcol}{blue!70!black}

\tikzset{>=latex} 
\tikzstyle{mass}=[line width=0.6,red!30!black,fill=red!40!black!10,rounded corners=1,
                  top color=red!40!black!20,bottom color=red!40!black!10,shading angle=20]
\tikzstyle{force}=[->,myred,very thick,line cap=round]
\tikzstyle{CM}=[red!40!black,fill=red!80!black!80]
\tikzstyle{rope}=[brown!70!black,very thick,line cap=round]
\tikzstyle{myarr}=[-{Latex[length=3,width=2]},thin]

\definecolor{darkblue}{rgb}{0,0,0.6}
\definecolor{darkred}{rgb}{0.6,0,0}

\def \stepSPMechanism {1.4}

\makeatletter
\renewcommand*{\fnum@figure}{{\normalfont \small{FIG.}~\thefigure}}
\makeatother

\makeatletter
\def\@seccntformat#1{\csname the#1\endcsname\quad}

\renewcommand\paragraph{\theparagraph.\arabic{paragraph}}
\makeatother

\newcommand{\beq}{\begin{equation}}
\newcommand{\eeq}{\end{equation}}
\newcommand{\ben}{\begin{equation*}}
\newcommand{\een}{\end{equation*}}
\newcommand{\bseq}{\begin{subequations}}
\newcommand{\eseq}{\end{subequations}}
\newcommand{\bea}{\begin{eqnarray}}
\newcommand{\eea}{\end{eqnarray}}
\newcommand{\bal}{\begin{align}}
\newcommand{\eal}{\end{align}}

\newcommand{\hb}{\boldsymbol{h}}

\newcommand{\hnb}{\hat{\boldsymbol{n}}}

\DeclareMathOperator{\Tr}{Tr}

\newcommand{\pb}[1]{\begingroup\color[rgb]{1,0,0}#1\endgroup}

\begin{document}

\title{Collective actuation in active solids in the presence of a polarizing field: \\a systematic analysis of the dynamical regimes}

\author{Paul Baconnier}
\affiliation{AMOLF, 1098 XG Amsterdam, The Netherlands.}
\affiliation{UMR CNRS Gulliver 7083, ESPCI Paris, PSL Research University, 75005 Paris, France.}
\author{Vincent Démery}
\affiliation{UMR CNRS Gulliver 7083, ESPCI Paris, PSL Research University, 75005 Paris, France.}
\affiliation{Univ Lyon, ENSL, CNRS, Laboratoire de Physique, F-69342 Lyon, France.}
\author{Olivier Dauchot}
\affiliation{UMR CNRS Gulliver 7083, ESPCI Paris, PSL Research University, 75005 Paris, France.}

\begin{abstract}
Collective actuation in active solids, the spontaneous condensation of the dynamics on a few elastic modes, takes place whenever the deformations of the structure reorient the forces exerted by the active units composing, or embedded in, the solid. In a companion paper, we show through a combination of model experiments, numerical simulations, and theoretical analysis that adding an external field that polarizes the active forces strongly affects the dynamical transition to collective actuation. A new oscillatory regime emerges, and a reentrance transition to collective actuation takes place. Depending on the degeneracy of the modes on which the dynamics condensates, and on the orientation of the field with respect to the stiff direction of the solid, several new dynamical regimes can be observed. The purpose of the present paper is to review these dynamical regimes in a comprehensive way, both for the single-particle dynamics and for the coarse-grained one. Whenever possible the dynamical regimes and the transition between them are described analytically, otherwise numerically.
\end{abstract}

\maketitle

\section{Introduction}

Active solids - dense assemblies or elastic structures composed of, or doped with, active units - encompass a wide class of systems ranging from biological to man-made materials \cite{koenderink2009active, menzel2013traveling, berthier2013non, prost2015active, briand2016crystallization, briand2018spontaneously, giavazzi2018flocking, woodhouse2018autonomous, ronceray2019stress, klongvessa2019active, maitra2019oriented, tan2022odd, Zheng2023, canavello2024polar, xi2024emergent, veenstra2025adaptive}, and exhibit exceptional mechanical properties \cite{scheibner2020odd, fruchart2023odd, veenstra2024non}. \textit{Collective actuation} describes self-sustained oscillations of such active solids, with a spontaneous condensation of the dynamics on a few vibrational modes of the elastic structure. This denomination was proposed in~\cite{baconnier2022}, where the authors describe and analyze the phenomenon in a model experimental system. The latter is composed of elementary polar particles, located at the nodes of an elastic lattice.
The orientation of these particles aligns with their motion, according to the so-called self-alignment mechanism, a now well-documented effect, introduced first in~\cite{Shimoyama1996}, and reintroduced independently in different contexts (see~\cite{baconnier2025self} for a recent review).
As a result, the displacements induced by the active forces reorients these forces, leading to a nonlinear elasto-active feedback.
When this feedback is strong enough, spontaneous oscillations take place, the dynamics condensates on essentially two modes of the elastic structure and collective actuation occurs. The phenomenon is however not limited to that model experimental system. It was clearly evidenced in large bacterial colonies~\cite{Xu2023}, and is likely to be present in other contexts, such as confined epithelial tissues~\cite{Peyret2019} and dense pedestrian crowds \cite{Gu2025}. More generally, confined assemblies of soft self-aligning polar particles are prone to exhibit collective actuation, as shown numerically in~\cite{Henkes2011}.

The simplicity of the experimental system proposed in~\cite{baconnier2022} allows for a description of the dynamics in terms of a set of overdamped Langevin equations, where the elasticity is described in the harmonic approximation, and the only nontrivial term, resulting from self-alignment, couples the orientational and translational degrees of freedom.
Taking advantage of this formulation, the authors could derive important results regarding the linear stability of the disordered phase, the selection of the modes by the dynamics, and, for some specific geometries, obtain a complete description of the transition, which we shall review below. This basic understanding of the mechanism of the transition further allowed to propose a way to switch between different regimes of collective actuation, controlling the mechanical tension inside the spring network~\cite{baconnier2023a}.
Also, whenever the solid hosts zero-energy deformation modes, stress propagation induces the spontaneous actuation of these modes, without exciting the finite energy vibrational ones.
In that case, the dynamics maps onto a relaxational dynamics in an effective Landau free energy, predicting mode selection and the onset of collective dynamics~\cite{hernandez2023}, as observed for instance in elastically connected swarms of robots~\cite{Ferrante2013a}.

Finally, the key role played by the coupling between the orientation of the active forces and the displacements, suggests the use of an external field to polarize these orientations and thereby manipulate the collective actuation dynamics, as simply as magnetic fields are used to manipulate spins. The ability of living systems to respond to various types of environmental cues is another motivation for analyzing the response of model systems to external fields \cite{sun2019infection, kennard2020osmolarity, goodwin2021mechanics, sengupta2021principles}.
This new avenue for controlling active solids is explored experimentally, numerically and theoretically in a companion paper~\cite{baconnier2025reentrant}. The variety of configurations that can be explored and the richness of the observed dynamics, however, call for a more systematic description.
This is the primary goal of the present work. 

In this paper, we will systematically analyze numerically and theoretically the dynamical regimes observed when varying the field amplitude and the strength of the elasto-active coupling. Depending on the network's structure and boundary conditions, which govern the normal mode spectrum of the passive structure, the two primary vibrational modes on which the dynamics condensates can be degenerate, or not \cite{baconnier2025reentrant}. Already in the absence of field, the latter case opens the way to a variety of periodic oscillations, as discussed in~\cite{Damascena2022} for the single-particle dynamics. When adding a polarizing field, we shall see that the orientation of the field with respect to the softest of these two modes also matters. The present analysis focuses on both the dynamics of a single self-aligning particle and that of the fields obtained from the coarse-graining of the N-particles microscopic model~\cite{baconnier2022}. The paper is organized as follows.
After an introduction of the microscopic model (Sec.~\ref{sec:micro}), the paper is divided into two sections, one dedicated to the single particle dynamics (Sec.~\ref{sec:single}) and one to the coarse-grained one (Sec.~\ref{sec:cg}).
Each section is divided according to the degeneracy or not of the modes of interest.

\section{Microscopic dynamics}
\label{sec:micro}

The active solids we consider are composed of a 2-dimensional elastic lattice, with $N$ nodes, at the location of which sits a polar active force of orientation $\hnb_i$, $i\in\{1,\dots N\}$, and amplitude $F_0$. Taken individually, this force extends or compresses a spring of stiffness $k$ by a length $l_e = F_0/k$.
The forces are exerted by self-aligning polar active particles, which reorient towards their velocity over a characteristic length $l_a$~\cite{baconnier2025self}.
The central control parameter of the dynamics is therefore the so-called elasto-active coupling $\Pi = l_e/l_a$. 
In the overdamped limit and harmonic approximation, the displacement $\boldsymbol{u}_{i}$ of the node $i$ with respect to the passive reference configuration and the orientation $\boldsymbol{n}_i$ follow
%
\begin{subequations} 
\label{eq:Npart_general}
\begin{align}
 \dot{\boldsymbol{u}}_{i} &= \Pi \boldsymbol{\hat{n}}_{i} - \mathbb{M}_{ij} \boldsymbol{u}_{j}, \label{eq1:Npart} \\
 \dot{\boldsymbol{n}}_{i} &= (\boldsymbol{\hat{n}}_{i} \times \left[ \dot{\boldsymbol{u}}_{i} + \boldsymbol{h} \right] ) \times \boldsymbol{\hat{n}}_{i} + \sqrt{2D} \eta_i,
 \label{eq2:Npart} 
\end{align}
\end{subequations}
where 
$\mathbb{M}$ is the dynamical matrix of the elastic lattice, $D$ sets the noise amplitude and $\eta_i$ are independent Gaussian white noises. Note that $(\mathbb{M}_{ i j })_{\beta \gamma}$ (resp. $(\mathbb{M}_{i i})_{\beta \gamma}$) encodes the strength of the elastic bond between nodes $i$ and $j$ (resp. the pinning constraints on node $i$), where $\beta, \gamma = \boldsymbol{\hat{e}}_x, \boldsymbol{\hat{e}}_y$ indicate the axis \cite{alexander1998amorphous, lubensky2015phonons}. Moreover, the above equations are made dimensionless, using $\gamma/k$ and $l_a$ as units of time and length, respectively, where $\gamma$ is an effective friction coefficient.
In the following, we shall only consider mechanically stable networks, thus all eigenvalues $\omega_k^2$ of the dynamical matrix $\mathbb{M}$ are positive. Note that we denote them as $\omega_{k}^{2}$ to reflect the fact that, in the case of passive underdamped dynamics, they correspond to the squared oscillation frequencies of the vibrational modes. In the present overdamped dynamics, they represent inverse relaxation times. In all cases, they are dimensionless quantities.
Finally $\hb$ is the external polarizing field, which can be added to act on the orientation of the active forces, as a magnetic field would do with XY spins.

The above model was studied in the zero-field case in~\cite{baconnier2022}.
The deterministic dynamics has a $N$-dimensional set of fixed points, where the active forces equilibrate with the elastic forces induced by the deformation:
any set of orientations defines one fixed point $\left(  \left\{ \boldsymbol{\hat{n}}_{i} \right\}, \left\{ \boldsymbol{u}_{i}  = \Pi  \mathbb{M}_{ij}^{-1} \boldsymbol{\hat{n}}_{j} \right\} \right)$).
The linear destabilization threshold $\Pi_c( \left\{ \boldsymbol{\hat{n}}_{i} \right\})$ depends on the fixed point configuration.
These thresholds are bounded $\Pi_c^{\text{min}}=\omega_\textrm{min}^2\le \Pi_c( \left\{ \boldsymbol{\hat{n}}_{i} \right\})  \le \Pi_c^{\text{max}}$, where $\omega_\textrm{min}^2$ is the smallest eigenvalue of the dynamical matrix $\mathbb{M}$.
For $\Pi<\omega_\textrm{min}^2$, all fixed points are marginally stable ($N$ zero eigenvalues) because of their rotational degeneracy.
$\Pi_c^{\text{max}}$ is not known analytically, but it is bounded from above by
\begin{equation}
\Pi^{\text{upp}} =\min_{\{i,j\}} \left(\frac{\omega^2_i +\omega^2_j}{c(| \boldsymbol{\varphi}_{i} \rangle,| \boldsymbol{\varphi}_{j} \rangle)}\right),
\label{eq:upperbound}
\end{equation}
where the function $c(\cdot,\cdot)$ only depends on the eigenvectors of $\mathbb{M}$, $\{ |\boldsymbol{\varphi}_{i} \rangle\}$. It is bounded between $0$ and $1$ and is maximal when the modes $|\boldsymbol{\varphi}_{i} \rangle$ and $|\boldsymbol{\varphi}_{j} \rangle$ are extended and locally orthogonal.
The collective actuation dynamics, where all the nodes oscillate in synchrony around their reference configuration takes place for $\Pi \ge \Pi_{\rm CA} \ge \omega_\textrm{min}^2$. Except for certain specific geometries, $\Pi_{\rm CA}$ could not be obtained analytically. Numerically, it is observed that  $\Pi_{\rm CA} \le \Pi_c^{\text{max}}$, leaving the place for a regime of coexistence between marginal fixed points and oscillating dynamics.

In the following, we shall concentrate on the effect of the external polarizing field on the transition to collective actuation. To do so we will make use of two simpler models derived from the one presented above. The first one is simply the one-particle version of it. The second one describes the coarse-grained dynamics of the displacement and polarization fields, obtained from a local averaging procedure of Eqs. (\ref{eq:Npart_general}), which we shall recall at the beginning of section~\ref{sec:cg}.


\section{Single particle dynamics}
\label{sec:single}

The single particle dynamics can be recast in:
\begin{subequations} 
\label{eq:1part_general}
\begin{align}
 \dot u_x &= \Pi \cos{\theta} - \omega_x^2 u_x, \label{eq1x:dimensionless_noiseless_braket} \\
 \dot u_y &= \Pi \sin{\theta} - \omega_y^2 u_y, \label{eq1y:dimensionless_noiseless_braket} \\
 \dot{\theta} &= -\sin{\theta}\, (\dot u_x + h_x) + \cos{\theta}\, (\dot u_y + h_y) + \sqrt{2D} \eta,
\end{align}
\end{subequations}
where $\theta$ denotes the orientation of $\boldsymbol{\hat{n}}=(\cos{\theta},\sin{\theta})$. There are two cases of interest, the degenerate one, $\omega^2_x = \omega^2_y = \omega^2_0$, and the non-degenerate one, for which we arbitrarily set $\omega^2_x < \omega^2_y$.

\subsection{The degenerate case: $\omega^2_x = \omega^2_y = \omega^2_0$}

As we shall see below, the rotational symmetry of the degenerate case in the absence of external field allows for a rather complete understanding of the phase diagram, that goes beyond the linear stability of the fixed points and captures the domain of existence of the nonlinear oscillating solutions too.

In the degenerate case, we define $\boldsymbol{u}=(R\cos\varphi,R\sin\varphi)$ and $\gamma=\theta-\varphi$.
With these notations, the deterministic version of  Eqs.~(\ref{eq:1part_general}) reads
\begin{subequations}
\label{eq:1part_degenerated}
\begin{align}
\dot{R} &= \Pi \cos \gamma - \omega_0^2 R, \\
\dot{\varphi} &= \frac{\Pi}{R} \sin \gamma, \\
\dot{\gamma} &= \left(\omega_0^2 R-\frac{\Pi}{R}\right) \sin\gamma - h \sin (\gamma+\varphi).
\end{align}
\end{subequations}

%

\begin{figure}[t!]
\centering
\begin{tikzpicture}

\node[] at (2.15,-0.3) {\includegraphics[height=3.0cm]{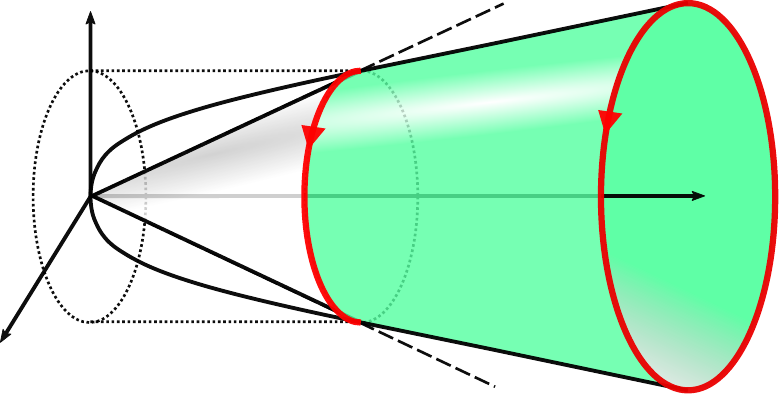}};

\node[rotate=10] at (3.0,0.67) {\small \textbf{CO}};
\node[rotate=24] at (1.0,0.00) {\small \textbf{FD}};

\node[] at (1.0,0.87) {\small $R = 1$};
\node[] at (-0.08,1.3) {\small $u_x$};
\node[] at (-0.93,-1.56) {\small $u_y$};
\node[] at (4.8,-0.3) {\small $\Pi$};
\node[] at (1.98,-1.61) {\small $\omega_{0}^{2}$};

\node[rotate=-24] at (1.0,-0.62) {\small $\Omega = 0$};
\node[rotate=-10] at (2.8,-1.25) {\small $\Omega > 0$};

\node[] at (-1.0,1.2) {\small (a)};


\filldraw[draw=black, fill=cyan, fill opacity=0.1] (-2.15, -5.9) -- (6.35,-5.9) -- (6.35,-2.5) -- (-2.15,-2.5) -- cycle;

\node[] at (-1.8,-2.15) {\small (b)};

\draw[->] (-1.2,-4.2) -- (5.8,-4.2);
\node[] at (6.0,-4.2) {\small $t$};

\node[rotate=90] at (-1.8,-3.35) {\small $\Pi > \omega_{0}^{2}$};
\node[rotate=90] at (-1.8,-5.0) {\small $\Pi < \omega_{0}^{2}$};

\tikzmath{ \xc = -0.8; \yc = -3.35; }
\tikzmath{ \xf = -0.8; \yf = -5.0; }

\tikzmath{ \xcc = \xc + \stepSPMechanism; \xccc = \xc + 2*\stepSPMechanism; \xcccc = \xc + 3*\stepSPMechanism; \xccccc = \xc + 4*\stepSPMechanism; \xcccccc = \xc + 5*\stepSPMechanism; \xccccccc = \xc + 6*\stepSPMechanism; \xcccccccc = \xc + 7*\stepSPMechanism; }
\tikzmath{ \xff = \xf + \stepSPMechanism; \xfff = \xf + 2*\stepSPMechanism; \xffff = \xf + 3*\stepSPMechanism; \xfffff = \xf + 4*\stepSPMechanism; \xffffff = \xf + 5*\stepSPMechanism; \xfffffff = \xf + 6*\stepSPMechanism; \xffffffff = \xf + 7*\stepSPMechanism; }

\tikzmath{ \angUa = -30; \angNa = 15; }
\tikzmath{ \angUb = -20; \angNb = 20; }
\tikzmath{ \angUc = -5; \angNc = 25; }
\tikzmath{ \angUd = 15; \angNd = 30; }
\tikzmath{ \angUe = 40; \angNe = 34; }
\tikzmath{ \angUf = 65; \angNf = 37; }
\tikzmath{ \angUg = 90; \angNg = 39; }
\tikzmath{ \angUh = 115; \angNh = 40; }

\tikzmath{ \cosUa = cos(\angUa); \sinUa = sin(\angUa); \cosUb = cos(\angUb); \sinUb = sin(\angUb); \cosUc = cos(\angUc); \sinUc = sin(\angUc); \cosUd = cos(\angUd); \sinUd = sin(\angUd); \cosUe = cos(\angUe); \sinUe = sin(\angUe); \cosUf = cos(\angUf); \sinUf = sin(\angUf); \cosUg = cos(\angUg); \sinUg = sin(\angUg); \cosUh = cos(\angUh); \sinUh = sin(\angUh); }
\tikzmath{ \cosNa = cos(\angUa+\angNa); \sinNa = sin(\angUa+\angNa); \cosNb = cos(\angUb+\angNb); \sinNb = sin(\angUb+\angNb); \cosNc = cos(\angUc+\angNc); \sinNc = sin(\angUc+\angNc); \cosNd = cos(\angUd+\angNd); \sinNd = sin(\angUd+\angNd); \cosNe = cos(\angUe+\angNe); \sinNe = sin(\angUe+\angNe); \cosNf = cos(\angUf+\angNf); \sinNf = sin(\angUf+\angNf); \cosNg = cos(\angUg+\angNg); \sinNg = sin(\angUg+\angNg); \cosNh = cos(\angUh+\angNh); \sinNh = sin(\angUh+\angNh); }

\tikzmath{ \angUi = -30; \angNi = 40; }
\tikzmath{ \angUj = -5; \angNj = 30; }
\tikzmath{ \angUk = 14; \angNk = 22; }
\tikzmath{ \angUl = 26; \angNl = 15; }
\tikzmath{ \angUm = 32; \angNm = 9; }
\tikzmath{ \angUn = 38; \angNn = 5; }
\tikzmath{ \angUo = 41; \angNo = 2; }
\tikzmath{ \angUp = 42; \angNp = 0; }

\tikzmath{ \cosUi = cos(\angUi); \sinUi = sin(\angUi); \cosUj = cos(\angUj); \sinUj = sin(\angUj); \cosUk = cos(\angUk); \sinUk = sin(\angUk); \cosUl = cos(\angUl); \sinUl = sin(\angUl); \cosUm = cos(\angUm); \sinUm = sin(\angUm); \cosUn = cos(\angUn); \sinUn = sin(\angUn); \cosUo = cos(\angUo); \sinUo = sin(\angUo); \cosUp = cos(\angUp); \sinUp = sin(\angUp); }
\tikzmath{ \cosNi = cos(\angUi+\angNi); \sinNi = sin(\angUi+\angNi); \cosNj = cos(\angUj+\angNj); \sinNj = sin(\angUj+\angNj); \cosNk = cos(\angUk+\angNk); \sinNk = sin(\angUk+\angNk); \cosNl = cos(\angUl+\angNl); \sinNl = sin(\angUl+\angNl); \cosNm = cos(\angUm+\angNm); \sinNm = sin(\angUm+\angNm); \cosNn = cos(\angUn+\angNn); \sinNn = sin(\angUn+\angNn); \cosNo = cos(\angUo+\angNo); \sinNo = sin(\angUo+\angNo); \cosNp = cos(\angUp+\angNp); \sinNp = sin(\angUp+\angNp); }

\filldraw[fill=none, draw=black] (\xc,\yc) circle (0.4);
\filldraw[fill=black] (\xc,\yc) circle (0.05);
\draw[densely dashed, thin, gray] (\xc,\yc) -- (\xc+0.8*\cosUa,\yc+0.8*\sinUa);
\draw[->, black, thick] (\xc,\yc) -- (\xc+0.4*\cosUa,\yc+0.4*\sinUa);
\draw[->, red, thick] (\xc+0.4*\cosUa,\yc+0.4*\sinUa) -- (\xc+0.4*\cosUa+0.4*\cosNa,\yc+0.4*\sinUa+0.4*\sinNa);

\filldraw[fill=none, draw=black] (\xcc,\yc) circle (0.4);
\filldraw[fill=black] (\xcc,\yc) circle (0.05);
\draw[densely dashed, thin, gray] (\xcc,\yc) -- (\xcc+0.8*\cosUb,\yc+0.8*\sinUb);
\draw[->, black, thick] (\xcc,\yc) -- (\xcc+0.4*\cosUb,\yc+0.4*\sinUb);
\draw[->, red, thick] (\xcc+0.4*\cosUb,\yc+0.4*\sinUb) -- (\xcc+0.4*\cosUb+0.4*\cosNb,\yc+0.4*\sinUb+0.4*\sinNb);

\filldraw[fill=none, draw=black] (\xccc,\yc) circle (0.4);
\filldraw[fill=black] (\xccc,\yc) circle (0.05);
\draw[densely dashed, thin, gray] (\xccc,\yc) -- (\xccc+0.8*\cosUc,\yc+0.8*\sinUc);
\draw[->, black, thick] (\xccc,\yc) -- (\xccc+0.4*\cosUc,\yc+0.4*\sinUc);
\draw[->, red, thick] (\xccc+0.4*\cosUc,\yc+0.4*\sinUc) -- (\xccc+0.4*\cosUc+0.4*\cosNc,\yc+0.4*\sinUc+0.4*\sinNc);

\filldraw[fill=none, draw=black] (\xcccc,\yc) circle (0.4);
\filldraw[fill=black] (\xcccc,\yc) circle (0.05);
\draw[densely dashed, thin, gray] (\xcccc,\yc) -- (\xcccc+0.8*\cosUd,\yc+0.8*\sinUd);
\draw[->, black, thick] (\xcccc,\yc) -- (\xcccc+0.4*\cosUd,\yc+0.4*\sinUd);
\draw[->, red, thick] (\xcccc+0.4*\cosUd,\yc+0.4*\sinUd) -- (\xcccc+0.4*\cosUd+0.4*\cosNd,\yc+0.4*\sinUd+0.4*\sinNd);

\filldraw[fill=none, draw=black] (\xccccc,\yc) circle (0.4);
\filldraw[fill=black] (\xccccc,\yc) circle (0.05);
\draw[densely dashed, thin, gray] (\xccccc,\yc) -- (\xccccc+0.8*\cosUe,\yc+0.8*\sinUe);
\draw[->, black, thick] (\xccccc,\yc) -- (\xccccc+0.4*\cosUe,\yc+0.4*\sinUe);
\draw[->, red, thick] (\xccccc+0.4*\cosUe,\yc+0.4*\sinUe) -- (\xccccc+0.4*\cosUe+0.4*\cosNe,\yc+0.4*\sinUe+0.4*\sinNe);




\filldraw[fill=none, draw=black] (\xf,\yf) circle (0.4);
\filldraw[fill=black] (\xf,\yf) circle (0.05);
\draw[densely dashed, thin, gray] (\xf,\yf) -- (\xf+0.8*\cosUi,\yf+0.8*\sinUi);
\draw[->, black, thick] (\xf,\yf) -- (\xf+0.4*\cosUi,\yf+0.4*\sinUi);
\draw[->, red, thick] (\xf+0.4*\cosUi,\yf+0.4*\sinUi) -- (\xf+0.4*\cosUi+0.4*\cosNi,\yf+0.4*\sinUi+0.4*\sinNi);

\filldraw[fill=none, draw=black] (\xff,\yf) circle (0.4);
\filldraw[fill=black] (\xff,\yf) circle (0.05);
\draw[densely dashed, thin, gray] (\xff,\yf) -- (\xff+0.8*\cosUj,\yf+0.8*\sinUj);
\draw[->, black, thick] (\xff,\yf) -- (\xff+0.4*\cosUj,\yf+0.4*\sinUj);
\draw[->, red, thick] (\xff+0.4*\cosUj,\yf+0.4*\sinUj) -- (\xff+0.4*\cosUj+0.4*\cosNj,\yf+0.4*\sinUj+0.4*\sinNj);

\filldraw[fill=none, draw=black] (\xfff,\yf) circle (0.4);
\filldraw[fill=black] (\xfff,\yf) circle (0.05);
\draw[densely dashed, thin, gray] (\xfff,\yf) -- (\xfff+0.8*\cosUk,\yf+0.8*\sinUk);
\draw[->, black, thick] (\xfff,\yf) -- (\xfff+0.4*\cosUk,\yf+0.4*\sinUk);
\draw[->, red, thick] (\xfff+0.4*\cosUk,\yf+0.4*\sinUk) -- (\xfff+0.4*\cosUk+0.4*\cosNk,\yf+0.4*\sinUk+0.4*\sinNk);

\filldraw[fill=none, draw=black] (\xffff,\yf) circle (0.4);
\filldraw[fill=black] (\xffff,\yf) circle (0.05);
\draw[densely dashed, thin, gray] (\xffff,\yf) -- (\xffff+0.8*\cosUl,\yf+0.8*\sinUl);
\draw[->, black, thick] (\xffff,\yf) -- (\xffff+0.4*\cosUl,\yf+0.4*\sinUl);
\draw[->, red, thick] (\xffff+0.4*\cosUl,\yf+0.4*\sinUl) -- (\xffff+0.4*\cosUl+0.4*\cosNl,\yf+0.4*\sinUl+0.4*\sinNl);

\filldraw[fill=none, draw=black] (\xfffff,\yf) circle (0.4);
\filldraw[fill=black] (\xfffff,\yf) circle (0.05);
\draw[densely dashed, thin, gray] (\xfffff,\yf) -- (\xfffff+0.8*\cosUm,\yf+0.8*\sinUm);
\draw[->, black, thick] (\xfffff,\yf) -- (\xfffff+0.4*\cosUm,\yf+0.4*\sinUm);
\draw[->, red, thick] (\xfffff+0.4*\cosUm,\yf+0.4*\sinUm) -- (\xfffff+0.4*\cosUm+0.4*\cosNm,\yf+0.4*\sinUm+0.4*\sinNm);




\end{tikzpicture}
\vspace{-0.15cm}
\caption{{\bf Single particle dynamics in a degenerate harmonic potential, in the absence of external field:} (a) The drift-pitchfork bifurcation; the white cone describes the set of marginal fixed points for increasing $\Pi$. For $\Pi=\omega_0^2$ all fixed points turn unstable and leave place to an orbiting solution, the oscillation frequency of which, $\Omega$ increases from zero at the transition.  (b) Destabilization mechanism; when $\Pi<\omega_0^2$, the displacement vector (black) catches up the orientation one (red) and the system restabilizes on a new fixed point; when $\Pi>\omega_0^2$, the displacement vector chases the orientation one indefinitely, leading to the oscillating solution.}
\label{fig:1part_nondege_nofield}
\end{figure}
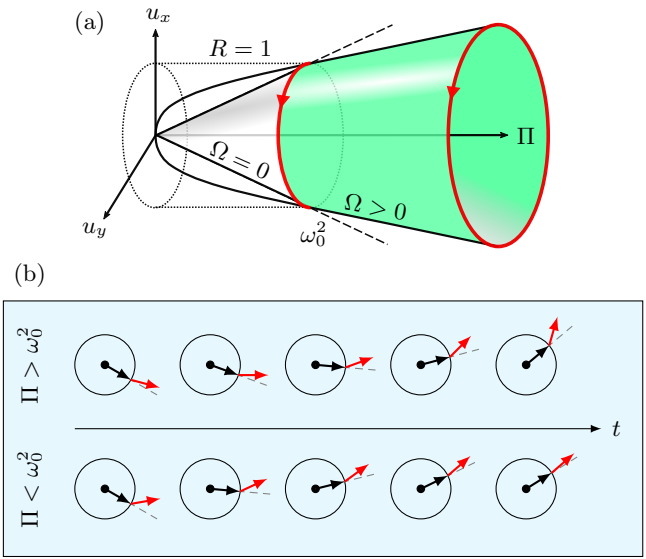

%


\subsubsection{In the absence of an external field \texorpdfstring{$\bm{h}=\bm{0}$}{h=0}}
\label{sec:single_h0}

When $h=0$ in Eqs.~(\ref{eq:1part_degenerated}), the system has an infinite set of fixed points $(R=\Pi/\omega_0^2$, $\varphi\in[0,2\pi]$, $\gamma=0)$.
The Jacobian of the linearized dynamics around any of the fixed point reads:
\begin{equation}
\begin{pmatrix}
-\omega_0^2 & 0 & 0 \\
0 &  0 & \omega_0^2 \\
0 &  0 & \Pi - \omega_{0}^{2} 
\end{pmatrix}.
\end{equation}
The dynamics along $R$, corresponding to the negative eigenvalue $-\omega_0^2$, always relaxes to the stationary value $R_0$. There is one zero eigenvalue associated with the rotational symmetry, and the last eigenvalue $\lambda = \Pi-\omega_0^2$ turns positive when $\Pi>\Pi_c = \omega_0^2$. 
All fixed points thus lose their stability for the same value of $\Pi$: they are marginally stable for $\Pi<\Pi_c$ and linearly unstable for $\Pi>\Pi_c$.
At the transition, the eigenvalues and the corresponding eigenvectors of the Jacobian coalesce, indicating that $(\Pi = \Pi_c, h=0)$ is an exceptional point~\cite{heiss2012physics}.

Beyond the instability threshold ($\Pi>\Pi_c$), there are periodic solutions with $R = (\Pi/\omega_0^2)^{1/2}$ and $\Omega = \dot\varphi=\pm\omega_0 ( \Pi - \omega_0^2)^{1/2}$~\cite{Dauchot2019}.
These Chiral Oscillations (CO) emerge continuously from the circular set of marginal fixed points at $\Pi = \Pi_c$ via a drift-pitchfork bifurcation (Fig.~\ref{fig:1part_nondege_nofield}-a and \cite{kness1992symmetry}). 
The physical mechanism behind this transition is that, when a small perturbation misaligns the displacement and the orientation vectors and $\Pi<\Pi_c$, the system re-stabilizes on a different fixed point. 
On the contrary, when $\Pi>\Pi_c$, the displacement vector cannot catch up to the orientation one, and the periodic dynamics sets in (Fig.~\ref{fig:1part_nondege_nofield}-b).

\subsubsection{Adding an external field \texorpdfstring{$\bm{h}$}{h}}
\label{sec:single_degenerate_h}

\begin{figure}[b!]
\centering
\begin{tikzpicture}

\node[rotate=0] at (0.0,0.0) {\includegraphics[width=0.99\columnwidth]{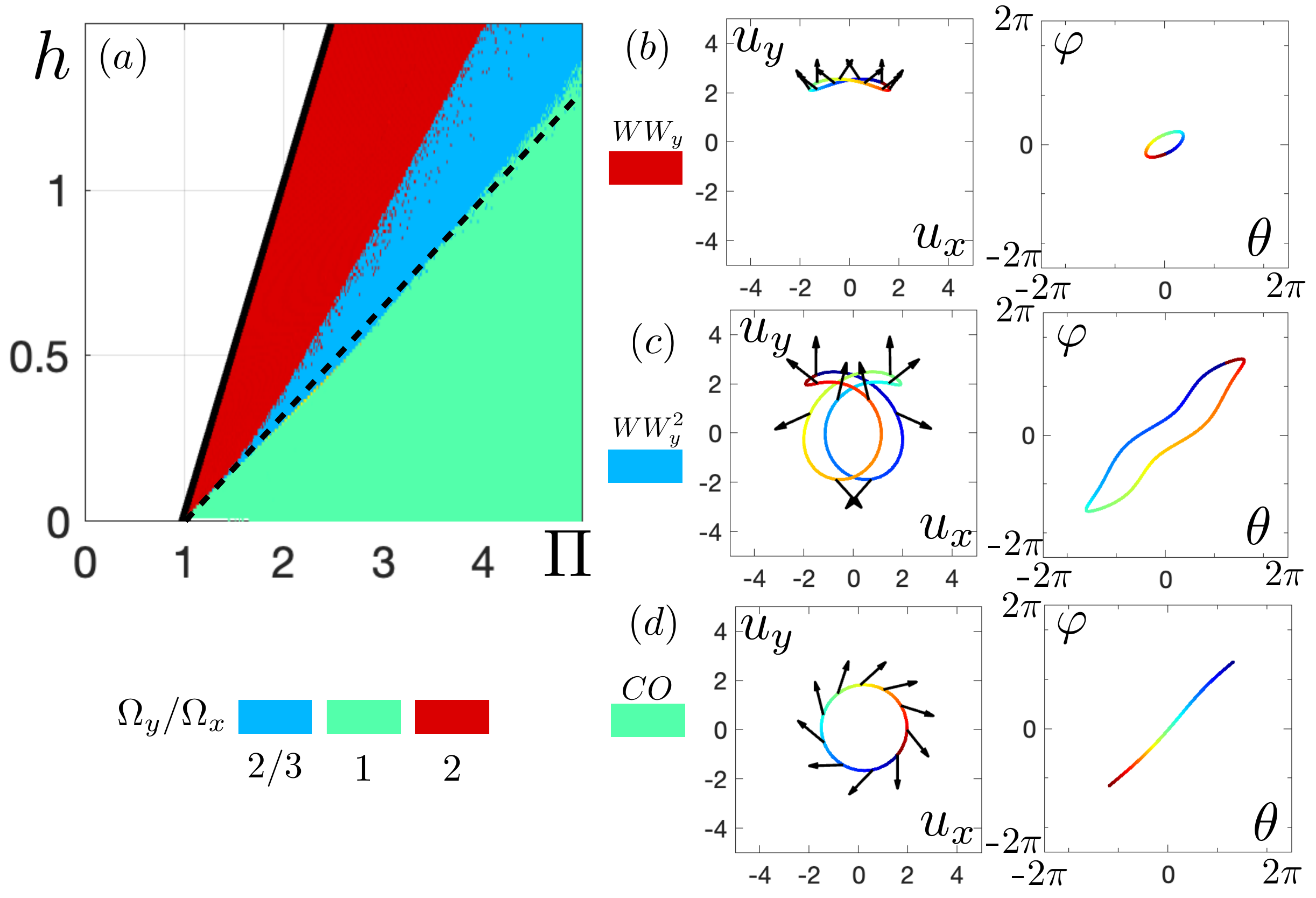}};

\node[rotate=0] at (-0.87,-0.25) {\small \textbf{CO}};

\node[rotate=73] at (-2.45,2.4) {\small \textbf{FP}};

\node[rotate=73] at (-2.03,2.22) {\small \textbf{WW}$_y$};

\node[rotate=52] at (-1.05,2.2) {\small \textbf{WW}$^2_y$};

\end{tikzpicture}
\vspace{-0.7cm}
\caption{{\bf Single particle dynamics in a degenerate harmonic potential, in the presence of an external field:} (a) Phase diagram; the color codes for the value of $\Omega_y/\Omega_x$ -- the ratio of the oscillation frequency of the displacement along the $x$ and $y$ directions -- as indicated in the legend (green : $\Omega_y/\Omega_x=1$, light blue : $\Omega_y/\Omega_x=2/3$, red: $\Omega_y/\Omega_x=2$); each point in the diagram is the result from an independent simulation, with random initial condition. The solid and dashed black lines indicate $\Pi_c = \omega_{0}^{2} + h$ and $\Pi^{\star} = \omega_{0}^{2} + 3h$, respectively. (b-c-d)-left: dynamics of the displacements in the WW$_y$, WW$_y^2$, and CO regimes; the trajectory is plotted during one period of oscillation, and colored with time running from dark blue to red; the dark arrows are snapshots of the orientation of the active force $\hnb$. (b-c-d)-right: corresponding dynamics of the phases $(\theta,\varphi)$ with the same color code.}
\label{fig:1part_dege_field}
\end{figure}
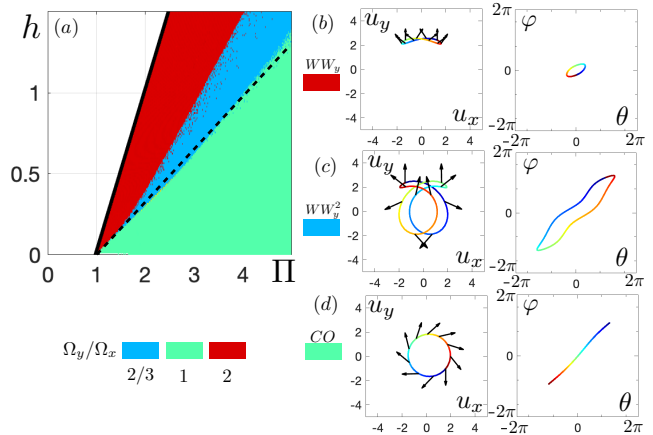

Adding a field breaks the rotational symmetry, responsible for the degeneracy of the fixed points. 
Figure~\ref{fig:1part_dege_field} displays the phase diagram obtained by solving Eqs.~(\ref{eq:1part_general}) numerically, for $\omega_0^2=1$, $\hb = h\hat{\bf e}_y$ and $D=0.1$.
For $\Pi < \omega_0^2$, the system is Frozen Polarized (FP), with the active force fluctuating around the direction of the polarizing field. For $\Pi > \omega_0^2$, three regimes are observed depending on the field amplitude and $\Pi$.
At small fields, the CO regime subsists, with a temporal modulation of the angle $\gamma$ at the CO rotation frequency (Fig.~\ref{fig:1part_dege_field}-d).
For intermediate fields, a new dynamical regime emerges, where the orientation of the active unit oscillates around that of the field, which translates in real space into a back-and-forth motion around the direction of the field, analogous to that of Windscreen Wiper (WW$_y$) (Fig.~\ref{fig:1part_dege_field}-b). Larger fields stabilize the FP state. 
For large enough $\Pi$, the CO and the WW$_y$ regimes are separated by a higher-order Windscreen Wiper regime, which we denote WW$_y^2$ (Fig.~\ref{fig:1part_dege_field}-c). These dynamical regimes can be distinguished by the bounded or unbounded nature of the phases $\theta$ and $\varphi$ as well as by $\Omega_y/\Omega_x$, the ratio of the oscillation frequency of the displacement along the $x$ and $y$ directions. While for the CO regime the phases are unbounded and $\Omega_y/\Omega_x=1$, in the WW regimes the phases are bounded and $\Omega_y/\Omega_x=2$, respectively $2/3$, for the WW$_y$, resp. the WW$_y^2$, regime. The transitions between the FP and the WW$_y$ regimes, respectively the WW$_y$ and the CO regimes, take place close to  $\Pi_c = \omega_0^2 + h$, and $\Pi^{\star} = \omega_0^2 + 3h$ (Fig.~\ref{fig:1part_dege_field}-a).

In the presence of an external field, the deterministic dynamics has only two fixed points, respectively polarized along, and opposite to, the direction of the field: $(R_0^{\pm}=\Pi/\omega_0^2,\gamma_0^{\pm}=0, \varphi_0^{\pm}=0 \ \text{or} \ \pi)$.
The fixed point pointing in the direction opposite to the field is always linearly unstable.
The one pointing in the direction of the field corresponds to the FP state.
The linear stability analysis indicates that the latter state destabilizes for $\Pi_c = \omega_0^2+h$, through a Hopf bifurcation with an oscillating frequency $\pb{\propto} \sqrt{h}$ at the transition.
The nature of the transition thus changes radically, as soon as an external field reduces the degeneracy of the fixed points.

\subsubsection{Expansion around the exceptional point}

We have shown that the three regimes FP, CO, and WW meet at the exceptional point $(\Pi=\Pi_c, h=0)$.
We now expand the dynamics around the exceptional point to get a quantitative insight into the nature of these regimes and the transitions between them.

We introduce the small parameter $\varepsilon = (\Pi - \omega_{0}^{2})/\omega_{0}^{2}$ and the rescaled field $\tilde{H} = h/(\varepsilon\omega_0^2)$, as suggested by the observed scaling of the transition lines $\Pi_c$ and $\Pi^{\star}$ (Sec.~\ref{sec:single_degenerate_h}).
Rescaling the variables as $R(t)=1+\epsilon\tilde\rho(\sqrt{\epsilon}t)$, $\varphi(t)=\tilde\varphi(\sqrt{\epsilon}t)$ and $\gamma(t)=\sqrt{\epsilon}\tilde\gamma(\sqrt{\epsilon}t)$, we find that the radius is a fast variable and that the dynamics of the angles reads, at order $\sqrt{\epsilon}$ (App.~\ref{app:pendulum}),
\begin{subequations} \label{eq:zeroth_order_pendulum}
\begin{align}
\dot{\tilde{\varphi}}  &= \tilde{\gamma}, \label{eq2:zeroth_order_pendulum} \\
\dot{\tilde{\gamma}} &= - \tilde{H} \sin \tilde{\varphi}+ \sqrt{\varepsilon} \tilde{\gamma} \left(1- \tilde{\gamma}^2 - \tilde{H} \cos \tilde\varphi \right). \label{eq3:zeroth_order_pendulum}
\end{align}
\end{subequations}

At order $\varepsilon^0$, the equations for $\tilde{\varphi}$ and $\tilde{\gamma}$ describe a weighing pendulum for the angle $\tilde{\varphi}$, suggesting a mapping of the bounded phase solution of the pendulum onto the WW$_y$ regime; and the unbounded phase solution of the pendulum onto the CO regime.
However, the energy of the pendulum, $E = \tilde{\gamma}^{2}/2 - \tilde{H} \cos \tilde{\varphi}$, defining the type of orbit,
is conserved, so that 
there is no mechanism to select the orbit at this order.

At order $\sqrt{\varepsilon}$, the second term in the right-hand side of Eq.~(\ref{eq3:zeroth_order_pendulum}) introduces an energy change $\delta E(\tilde H, E)\propto\sqrt{\varepsilon}$ over a period of the pendulum.
Equilibrium orbits satisfy $\delta E = 0$, and stable ones require $\partial \delta E/\partial E < 0$. 
We can compute $\delta E$ in the limiting cases of large energy, small amplitudes, and for the heteroclinic orbit $E = \tilde{H}$ (App.~\ref{app:pendulum}):
\begin{subequations}
\begin{align}
  \delta E\left(\tilde H, E\gg \{\tilde H, 1\}\right) & = -4\pi\sqrt{2\varepsilon}E^{3/2},\\
  \delta E\left(\tilde H, E\simeq -\tilde H\right) & = 2\pi\sqrt{\frac{\varepsilon}{\tilde H}}(1-\tilde H)(E+\tilde H),\label{eq:de_small}\\
  \delta E\left(\tilde H, E=\tilde H\right) & = 8 \sqrt{\varepsilon\tilde{H}} (1 - 3\tilde{H}).\label{eq:de_hetero}
\end{align}
\end{subequations}

We recover the stability range of the FP regime obtained with the linear stability analysis from Eq.~(\ref{eq:de_small}): the minimum energy state is stable if $\tilde H>1$ ($\delta E<0$), while it is unstable if $\tilde H<1$.
In the latter case, the system finds one of the two oscillating regimes, WW$_y$ or CO.
From equation (\ref{eq:de_hetero}), we see that, if $\tilde{H} > 1/3$, the energy decays on the heteroclinic orbit, leading to the selection of the WW$_y$ regime. 
In contrast, if $\tilde{H} < 1/3$, the energy increases on the heteroclinic orbit, corresponding to the selection of the CO regime. 
We thus recover the  scaling observed numerically for $\Pi^*(h)$.

\begin{figure*}[t!]
\centering

\begin{tikzpicture}

\node[rotate=0] at (-6.2,-5.6) {\includegraphics[width=5.9cm]{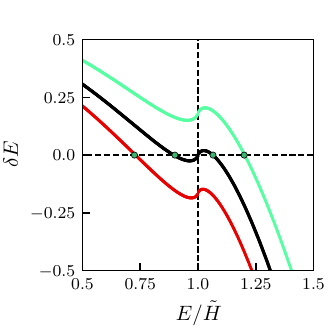}};

\node[rotate=-39] at (-7.0,-4.42) {\footnotesize $\tilde{H} = 1/3$};
\node[rotate=-34] at (-7.0,-3.85) {\footnotesize \color{colorCO_new} $\tilde{H} = 0.32$};
\node[rotate=-44] at (-7.0,-5.48) {\footnotesize \color{colorWW_new} $\tilde{H} = 0.345$};

\node[rotate=0] at (-0.5,-5.6) {\includegraphics[width=5.9cm]{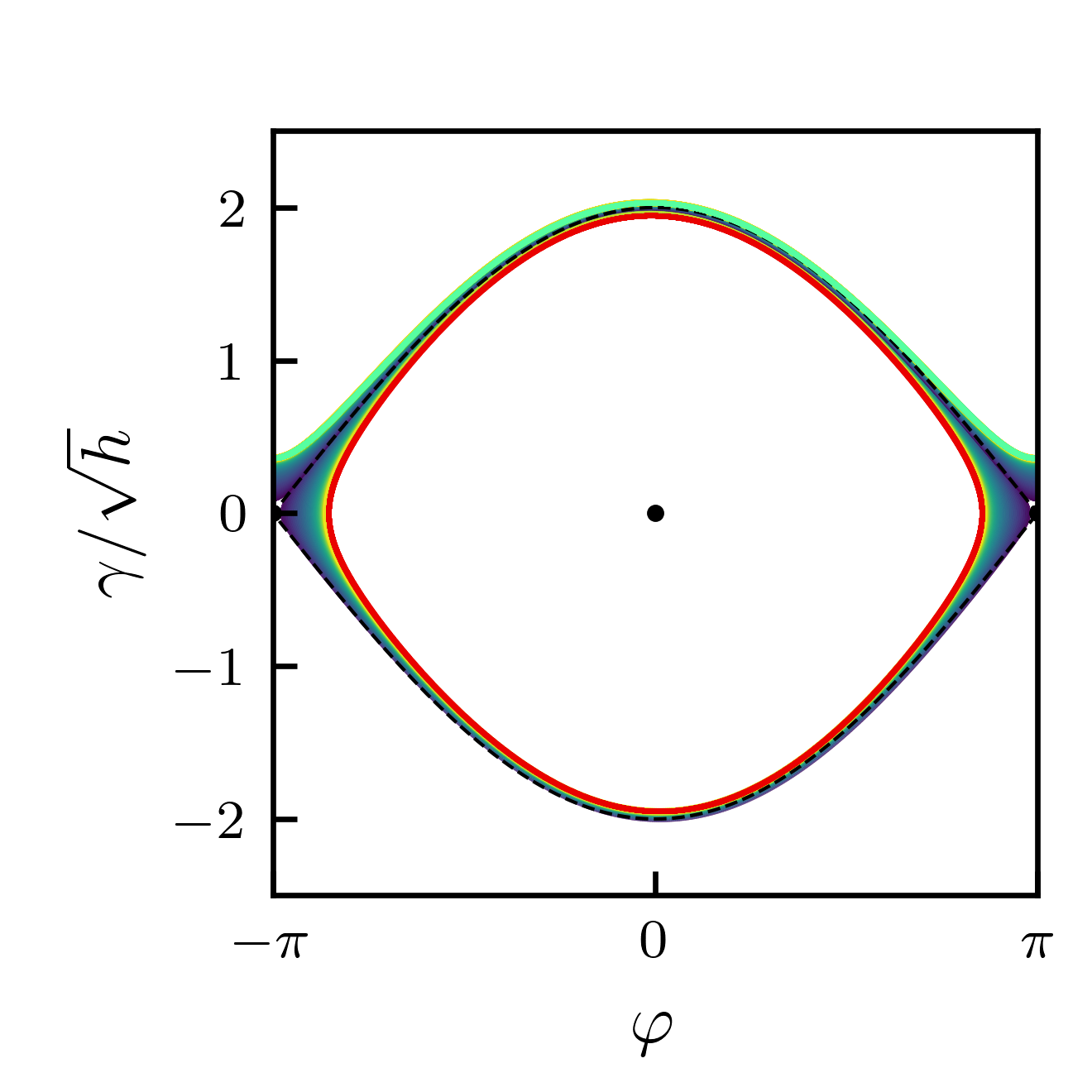}};

\node[rotate=0] at (1.4,-3.65) {\small $\tilde{H} = 1/3$};

\node[rotate=44] at (-1.22,-4.15) {\small \color{colorCO_new} CO};
\node[rotate=44] at (-0.9,-4.65) {\small \color{colorWW_new} WW$_y$};

\node[rotate=0] at (5.3,-5.6) {\includegraphics[width=5.9cm]{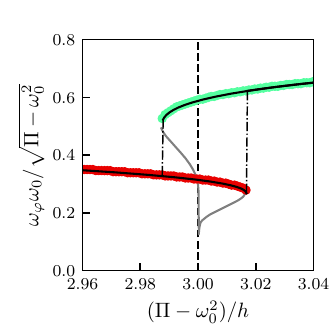}};

\draw[->, thick, color=black] (5.0,-5.05) -- (5.0,-5.55);
\draw[->, thick, color=black] (7.0,-5.55) -- (7.0,-5.05);

\node[rotate=-4] at (4.45,-6.05) {\small \color{colorWW_new} WW$_y$};
\node[rotate=10] at (7.2,-3.9) {\small \color{colorCO_new} CO};

\node[rotate=0] at (-3.9,-3.68) {\small (a)};
\node[rotate=0] at (-1.63,-3.68) {\small (b)};
\node[rotate=0] at (4.16,-3.68) {\small (c)};

\end{tikzpicture}

\vspace*{-0.4cm}
\caption{\small{\textbf{Single particle in a degenerate harmonic potential: mapping with the physical pendulum close to the exceptional point}.
(a) Energy drift $\delta E$ as a function of $E/\tilde{H}$, for different values of $\tilde{H}$ (App.~\ref{app:pendulum}).
Stable orbits are highlighted with a green marker.
(b) Phase portrait of two transient regimes at $\tilde{H} = 1/3$, with initial energies slightly above and below the heteroclinic orbit's energy (black dashed line); the heteroclinic orbit is the unstable fixed point in panel (a) for $\tilde{H} = 1/3$, and the separatrix of transients.
The stationary regimes obtained correspond to the stable orbits shown in panels (a) for $\tilde{H} = 1/3$.
(c) Rescaled fundamental frequency of oscillation of $\varphi$ as a function of the rescaled distance to the threshold, for a small field $H = 10^{-4}$. Colored markers represent numerical simulations (green: CO, red: WW$_y$), and the solid black lines (resp. solid grey line) are the stable (resp. unstable) solutions of the pendulum equations.
}}
\label{fig:mapping_pendulum}
 \vspace*{-0.3cm}
\end{figure*}
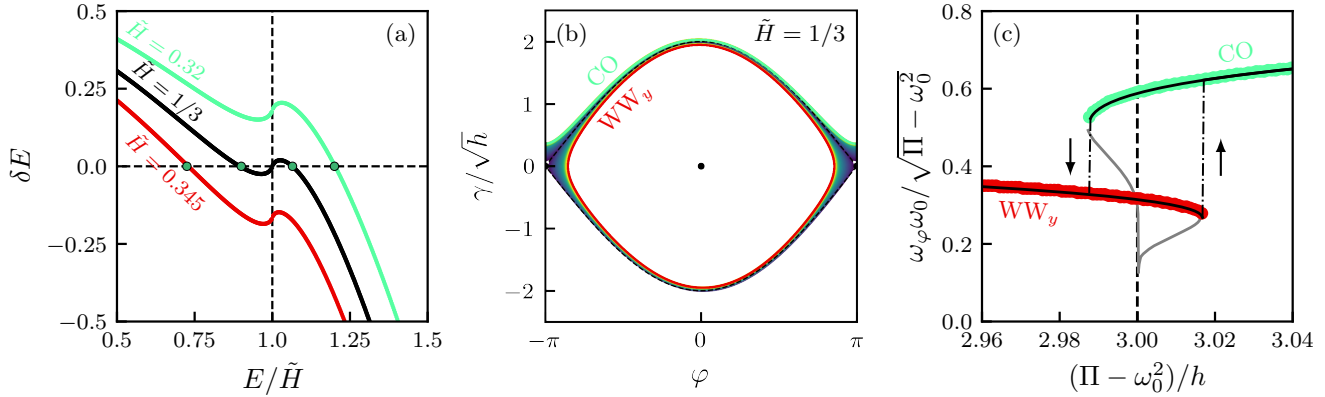

We can compute the energy change $\delta E(\tilde{H},E)$ numerically (Fig. \ref{fig:mapping_pendulum}-a).
For $\tilde{H} = 1/3$, we actually find two stable solutions, one bounded ($E < \tilde{H}$) and one unbounded ($E > \tilde{H}$).
This is the hallmark of a hysteresis: two stable solutions coexist within a small range of $\tilde{H}$. 
For a large enough external field $\tilde{H} > \tilde{H}_{+} \simeq 0.3357$, the only stable solution is bounded, corresponding to a WW$_y$ regime.
For a small enough external field $\tilde{H} < \tilde{H}_{-} \simeq 0.3314$, the only stable solution is unbounded, corresponding to a CO regime.
In contrast, within the range $[\tilde{H}_{-}, \tilde{H}_{+}]$, the initial condition sets the stationary solution reached by the system.
Remarkably, the transition from regime WW$_y$ to regime CO is a double saddle-node bifurcation of limit cycles (Fig. \ref{fig:mapping_pendulum}-c).
Indeed, as one stable limit cycle reaches instability, it collides with an unstable limit cycle, which exists in the entire coexistence region, and goes through the zero-frequency heteroclinic orbit at precisely $\tilde{H} = 1/3$.

Finally, we simulate Eqs.~(\ref{eq:1part_degenerated}), placing ourselves at $\tilde{H} = 1/3$, very close to the exceptional point ($H = 10^{-4}$).
We start from two initial conditions, which, within the above mapping, have energies $E = \tilde{H} \pm \delta$, where $0 < \delta \ll 1$, i.e. slightly below and slightly above the heteroclinic orbit's energy.
We find that both initial conditions converge toward the predicted stable orbits (Fig. \ref{fig:mapping_pendulum}-b).
The transient regime is very long, as expected close to the exceptional point. 
After confirming the hysteresis, we perform an annealing simulation, slowly varying $\tilde{H}$ back-and-forth around $1/3$, keeping $H = 10^{-4}$. 
We compare the main frequency of oscillation of $\varphi$ (obtained from the largest peak of its FFT) to the frequency of the stable pendulum solutions found above. 
We find a perfect agreement between the predictions and the numerical data (Fig. \ref{fig:mapping_pendulum}-c), with a hysteresis loop.
Note that this coexistence is not visible in Fig.~\ref{fig:1part_dege_field}a because of its minute range.
Interestingly, as one approaches the transitions, the WW$_y$ and CO regimes' frequency never vanishes, meaning that no stable pendulum orbits are selected in the vicinity of the heteroclinic orbit.


\subsubsection{Expansion close to the FP-WW transition}

To understand better the transition between the FP and WW regimes, we expand the dynamics around this transition by setting $\Pi = \Pi_c + \varepsilon$, with $\varepsilon \ll 1$.
There, the amplitude of the weakly non-linear limit cycle branching off from the FP regime, corresponding to the WW$_y$ regime, can be computed using multiple-scale analysis (App.~\ref{app:multiple_scale_ww}).

One introduces the slow timescale $T=\varepsilon t$ and expands the variables $R(t), \varphi(t), \theta(t)$ in $\varepsilon$.
At order $0$, only the equation on $R$ is nontrivial.
At the order $1/2$, we find for $\varphi$ and $\theta$,
\begin{equation}
\begin{pmatrix}
\varphi \\
\theta 
\end{pmatrix} = \sqrt{\varepsilon}A(T)
\begin{pmatrix}
1 \\
1 + i\sqrt{h} 
\end{pmatrix}  e^{i\sqrt{h}t} + \text{c.c.},  
\end{equation}
where c.c. indicates the complex conjugate of the previous term. Solving for orders $1$ and $3/2$, we find the amplitude equation
\begin{equation}
\dot A = \frac{A}{2} - Z \left|A\right|^2A,
\end{equation}
where $Z$ is a complex number (App.~\ref{app:multiple_scale_ww}), from which we deduce the amplitude
\begin{equation} \label{eq:final_result_WW_weakly_nonlinear}
 |A| = \frac{1}{\sqrt{2 Z_r}} = \sqrt{ \frac{ 2(4h+1) }{ h(h+1)(8h + 5) } },
\end{equation}
where $Z_r$ is the real part of $Z$.
The amplitudes of the oscillations of $\varphi$ and $\theta$ are proportional to $\sqrt{\varepsilon}$, indicating a supercritical Hopf bifurcation.

\subsubsection{Small field expansion}

For $h=\delta h$, with $\delta h \ll 1$, and far enough from the exceptional point, one can linearize the dynamics around the stationary CO regime  ($R_0 = \sqrt{\Pi}/\omega_{0}$, $\cos \gamma_0 = \omega_0/\sqrt{\Pi}$, $\dot{\varphi} = \Omega_0 = \omega_0\sqrt{\Pi - \omega_{0}^{2}}$) (App~\ref{app:linear_response_CO}). 
Introducing the small quantities $R(t) = R_0 + \delta R(t)$, $\gamma(t) = \gamma_0 + \delta \gamma(t)$, and $\varphi(t) = \Omega_0 t + \delta \varphi(t)$, we show that the CO regime is linearly stable and that the field acts as a sinusoidal forcing of amplitude $\delta h$ and frequency $\Omega_0$.
The amplitude of the resulting oscillations $(\delta R, \delta \varphi, \delta \gamma) = (A_R, A_{\varphi}, A_{\gamma}) e^{i\Omega_0 t}$ depends on the distance to the exceptional point, with
\begin{equation} \label{eq:LR_CO_ansatz_sol}
\begin{pmatrix}
 | A_{R} | \\
 | A_{\varphi} | \\
 | A_{\gamma} |
\end{pmatrix} \simeq
\delta h \begin{pmatrix}
1/\omega_{0}^{2} \\
1/ (\Pi - \omega_{0}^{2}) \\
1/(\omega_{0}\sqrt{\Pi - \omega_{0}^{2}})
\end{pmatrix}.
\end{equation}
The divergence of the modulation amplitudes along $\varphi$ and $\gamma$ when $\Pi\rightarrow \omega_0^2$ points at a singular behavior at the exceptional point.


\subsection{The non-degenerate case: $\omega^2_x < \omega^2_y$}

In the non-degenerate case, the rotational symmetry is broken, even in the absence of an external field.
As we shall see, this significantly modifies the phase diagram, but also reduces our ability to make precise analytical statements beyond the linear stability of the fixed points.
Also, the orientation of the field with respect to the stiff $(y)$, or soft $(x)$, direction matters.

\subsubsection{In the absence of an external field \texorpdfstring{$\bm{h}=\bm{0}$}{h=0}}

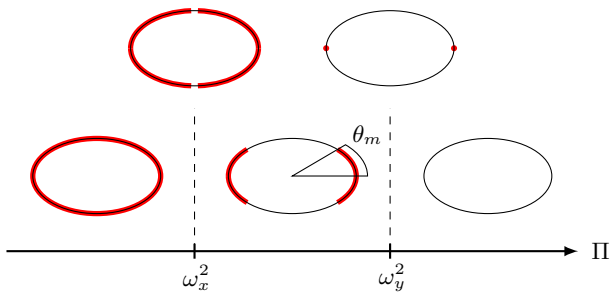
\begin{figure}[b!]
\centering

\begin{tikzpicture}


\draw[->, thick] (-3.8,-5.0) -- (3.8,-5.0);
\node[rotate=0] at (4.1,-5.0) {\small $\Pi$};

\draw[thick] (-1.3,-5.1) -- (-1.3,-4.9);
\draw[thick] (1.3,-5.1) -- (1.3,-4.9);

\node[rotate=0] at (1.33,-5.35) {\small $\omega_{y}^{2}$};
\node[rotate=0] at (-1.27,-5.35) {\small $\omega_{x}^{2}$};

\draw[dashed] (-1.3,-5.0) -- (-1.3,-3.1);
\draw[dashed] (1.3,-5.0) -- (1.3,-3.1);

\draw[red, line width=0.07cm] (0.85,-4.0) arc[start angle=0, end angle=45, x radius=0.85, y radius=0.5];
\draw[red, line width=0.07cm] (0.85,-4.0) arc[start angle=0, end angle=-45, x radius=0.85, y radius=0.5];

\draw[red, line width=0.07cm] (-0.85,-4.0) arc[start angle=180, end angle=225, x radius=0.85, y radius=0.5];
\draw[red, line width=0.07cm] (-0.85,-4.0) arc[start angle=180, end angle=135, x radius=0.85, y radius=0.5];

\draw[] (0.0,-4.0) ellipse (0.85 and 0.5);

\draw[] (0.0,-4.0) -- (1.0,-4.0)
        -- +(0.0,0.0) arc[start angle=0, end angle=45, x radius=1.0cm, y radius=0.588cm] -- cycle;

\draw[] (2.6,-4.0) ellipse (0.85 and 0.5);

\draw[red, line width=0.07cm] (-2.6,-4.0) ellipse (0.85 and 0.5);
\draw[] (-2.6,-4.0) ellipse (0.85 and 0.5);

\fill[red] (0.45,-2.3) circle (0.04);
\fill[red] (2.15,-2.3) circle (0.04);
\draw[] (1.3,-2.3) ellipse (0.85 and 0.5);

\draw[red, line width=0.07cm] (-0.45,-2.3) arc[start angle=0, end angle=87, x radius=0.85, y radius=0.5];
\draw[red, line width=0.07cm] (-0.45,-2.3) arc[start angle=0, end angle=-87, x radius=0.85, y radius=0.5];

\draw[red, line width=0.07cm] (-2.15,-2.3) arc[start angle=180, end angle=93, x radius=0.85, y radius=0.5];
\draw[red, line width=0.07cm] (-2.15,-2.3) arc[start angle=180, end angle=267, x radius=0.85, y radius=0.5];

\draw[] (-1.3,-2.3) ellipse (0.85 and 0.5);

\node[rotate=0] at (1.0,-3.45) {\small $\theta_m$};

\end{tikzpicture}
\vspace{-0.35cm}
\caption{{\bf Linear stability of the fixed points of the dynamics for a single particle in a non-degenerate harmonic potential with  $\omega^2_x < \omega^2_y$, in the absence of an external field:} The fixed points, represented in the ($u_x$,$u_y$)-plane, are distributed along an ellipse, as given by Eq. (\ref{eq:ellipse}), and the stability threshold $\Pi_c$ depends on the orientation $\theta_0$ (which is also the angle with respect to the $x$-axis of this fixed point).
The red overlay indicates the marginally stable fixed points; when absent the fixed points are linearly unstable.}
\label{fig:1part_nondege_nofield_fp}
\end{figure}

Coming back to Eqs. (\ref{eq:1part_general}), we consider the fixed points and their stability threshold $\Pi_c$. There is again an infinite set of fixed points parametrized by the orientation $\theta_0$, $(\theta_0,\boldsymbol{u}  = (\Pi \cos\theta_0/\omega_x^2, \Pi \sin\theta_0/\omega_y^2))$, which now describe an ellipse of equation
\begin{equation} \label{eq:ellipse}
\omega_x^4 u_x^2 + \omega_y^4 u_y^2 = \Pi^2.
\end{equation}
Here, the linear stability of the fixed points depends on the orientation $\theta_0$: $\Pi_c(\theta_0)= \omega_x^2 \omega_y^2 / (\omega_x^2 \cos\theta_0 + \omega_y^2 \sin\theta_0)$; so that for $\Pi > \Pi_c$, the fixed points oriented along $\theta_0$ are linearly unstable (Fig.~\ref{fig:1part_nondege_nofield_fp}). There is again always a zero eigenvalue, so that even stable fixed points are marginal.
The fixed points which destabilize at the smallest value of $\Pi$ are the ones where the particle points in the stiffest direction $(\theta_0 = \pm \pi/2)$, with $\Pi_c = \omega_x^2$, leaving two disconnected sets of marginally stable fixed points for $\omega_{x}^{2} < \Pi < \omega_{y}^{2}$ (Fig.~\ref{fig:1part_nondege_nofield_fp}-left).
The fixed points which destabilize at the largest value of $\Pi$ are the ones where the particle points in the softest direction $(\theta_0 = 0,\pi)$, with $\Pi_c = \omega_y^2$. For $\Pi > \omega_y^2$, all fixed points are unstable (Fig.~\ref{fig:1part_nondege_nofield_fp}-right).

At large enough activity, no fixed points are stable, thus a dynamical regime must set in.
The periodic solutions have been studied in Ref.~\cite{Damascena2022}, revealing a rich behavior depending on $\beta=\Pi/\omega_x^2$ and the eccentricity of the harmonic potential, $\epsilon = (\omega_y^2- \omega_x^2)/\omega_x^2$ (see Fig.~2 of Ref.~\cite{Damascena2022}). 
For small enough eccentricity $\epsilon$, the authors recover elliptical chiral orbits, similar to the degenerate case studied above (Sec. \ref{sec:single_h0}); while for larger $\epsilon$, many different periodic solutions are observed.
The orbits can be classified according to the number $p$ of self-crossings performed by the particle trajectory into the $(x,y)$ plane: $p = 0$ for elliptic orbits (such as the circular chiral orbits of the degenerate case),  $p = 1$ for lemniscates (figure-eight-like motion), and $p > 1 $ for higher-order generalized lemniscates.
In the same vein, there are two types of periodic motion: the phase-unbounded one, meaning that $\boldsymbol{\hat{n}}$ performs a full $2\pi$ rotation ({\bf R}) within one period, either clockwise or counterclockwise; and the phase-bounded ones, where $\boldsymbol{\hat{n}}$ swings back and forth in librational ({\bf L}) motion.
Orbits with even $p$ (resp. odd $p$) are of type {\bf R} (resp. {\bf L}).
The phase diagram obtained numerically also indicates that as soon as $\omega_x^2<\omega_y^2$ ($\epsilon>0$), there are orbiting solutions existing in the range $\omega_x^2<\Pi<\omega_y^2$, coexisting with the sets of marginal fixed points.
However, the range of coexistence seems to decrease with increasing ellipticity $\epsilon$.

The role of the noise on the dynamics has not been studied in detail and only numerical results are available.
The orientation of the particle diffuses on the continuous sets of marginal fixed points and the noise blurs the orbiting dynamics.
For very large noise, one observes a Boltzmann-like density centered on the mechanical equilibrium.
There is however a case of interest that was described in Ref.~\cite{baconnier2024noise}, when $\omega_x^2\ll \omega_y^2$ and $\omega_x^2\le\Pi\ll\omega_y^2$.
In this case, the deterministic dynamics is frozen on one of the marginal fixed points set by the initial condition.
When adding noise, one expects the dynamics to diffuse in each continuous set of marginal fixed points, with the possibility of stochastic jumps between the two sets.
Numerical simulations indeed report such jumps, however,
the analysis of the power spectrum of the dynamics reveals that they take place with a temporal regularity.
So far, no proper stochastic theory accounts for this peculiar dynamics.

At the collective level, this oscillating dynamics translates into a so-called Noise-Induced Collective Actuation (NICA) regime, whose amplitude and frequency increase with $\Pi$ and the noise amplitude \cite{baconnier2024noise}. The emergence of such oscillations is well captured by the coarse-grained dynamics, as described below.


\begin{figure*}[t!]
\centering

\begin{tikzpicture}

\node[rotate=0] at (0.0,0.0) {\includegraphics[width=0.95\textwidth]{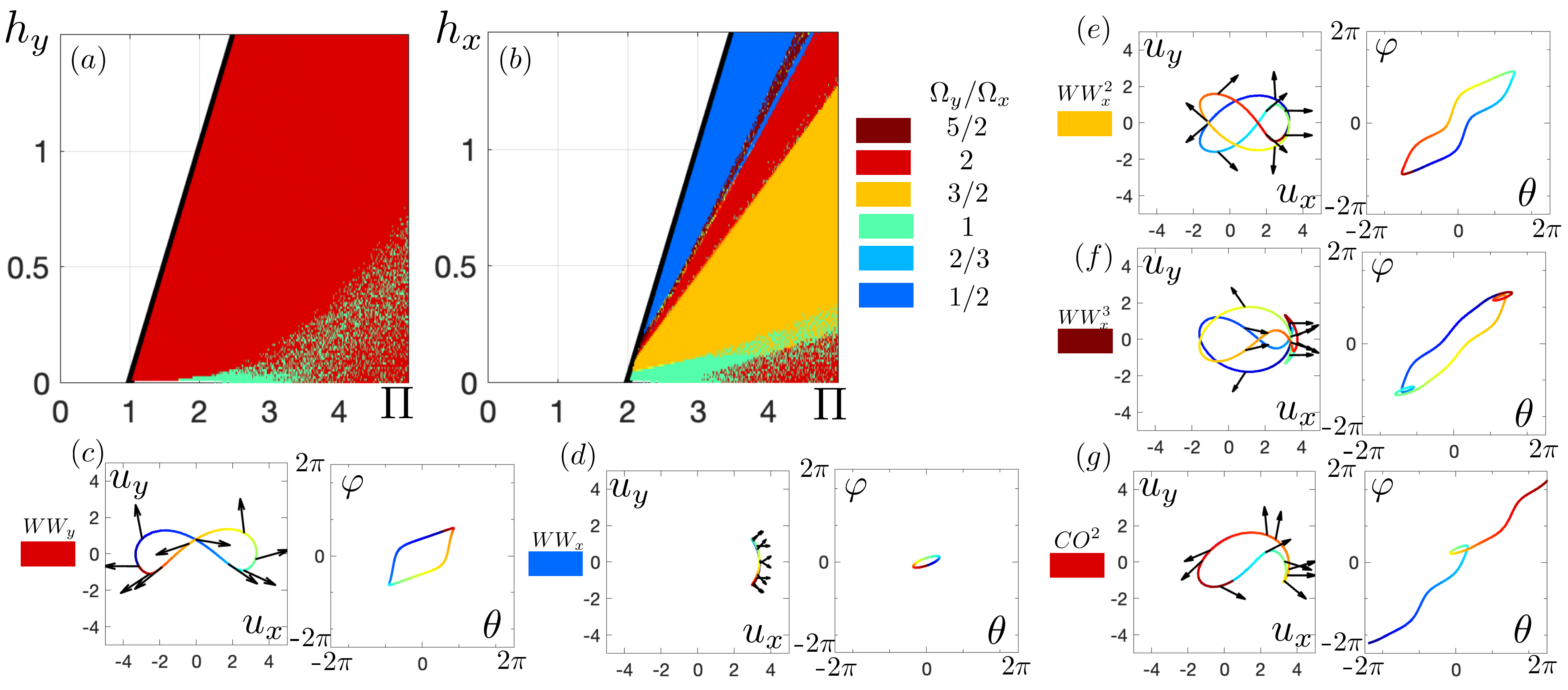}};

\node[rotate=73] at (-6.35,2.95) {\small \textbf{FP}};
\node[rotate=73] at (-5.86,2.81) {\small \textbf{WW}$_y$};

\node[] at (-5.65,0.46) {\footnotesize \textbf{CO}};
\draw[] (-5.95,0.28) -- (-5.35,0.28);
\draw[->] (-5.65,0.28) -- (-5.85,-0.34);

\node[rotate=73] at (-0.95,2.95) {\small \textbf{FP}};
\node[rotate=73] at (-0.46,2.81) {\small \textbf{WW}$_x$};

\node[rotate=52] at (0.1,1.70) {\small \textbf{WW}$^2_x$};

\node[] at (-0.85,0.46) {\footnotesize \textbf{CO}};
\draw[] (-1.15,0.28) -- (-0.55,0.28);
\draw[->] (-0.85,0.28) -- (-1.15,-0.34);

\node[] at (0.05,0.66) {\footnotesize \textbf{WW}$_y$};
\draw[] (-0.25,0.48) -- (0.35,0.48);
\draw[->] (0.05,0.48) -- (0.3,-0.34);

\node[] at (1.05,3.36) {\footnotesize \textbf{CO}$^2$};
\draw[] (0.75,3.18) -- (1.35,3.18);
\draw[->] (1.05,3.18) -- (0.35,2.76);

\end{tikzpicture}
\vspace{-0.35cm}
\caption{{\bf Single particle dynamics in a non-degenerate harmonic potential $(\omega_x^2 = 1, \omega_y^2 = 2)$ , in the presence of an external field:} (a) Phase diagram with the field in the stiff direction $\boldsymbol{h} = h_y \boldsymbol{\hat{e}}_y$; (b) phase diagram with the field in the soft direction $\boldsymbol{h} = h_x \boldsymbol{\hat{e}}_x$; the color codes for the value of $\Omega_y/\Omega_x$ -- the ratio of the oscillation frequency of the displacement along the $x$ and $y$ directions -- as indicated in the legend. Each point in the diagram is the result of an independent simulation, with a random initial condition. As such, areas of parameter space where two colors appear intermingled, indicate zones of coexistence between different dynamical regimes. The solid black lines indicate $\Pi_c(h_{x/y})$, as obtained from linear stability analysis. (c-g)-left: dynamics of the displacements in the dynamical regimes as named; the trajectory is plotted during one period of oscillation, and colored with time running from dark blue to red; the dark arrows are snapshots of the orientation of the active force $\hnb$. (c-g)-right: corresponding dynamics of the phases $(\theta,\varphi)$ with the same color code. Nota bene: the WW$_y$ and the CO$^2$ regimes share the same color code because $\Omega_y/\Omega_x=2$ for both of them; they can however not be confused because WW$_y$ is a bounded phases regime, while the phases are unbounded in the CO$^2$ regime.}
\label{fig:1part_nondege_wfield}
\end{figure*}
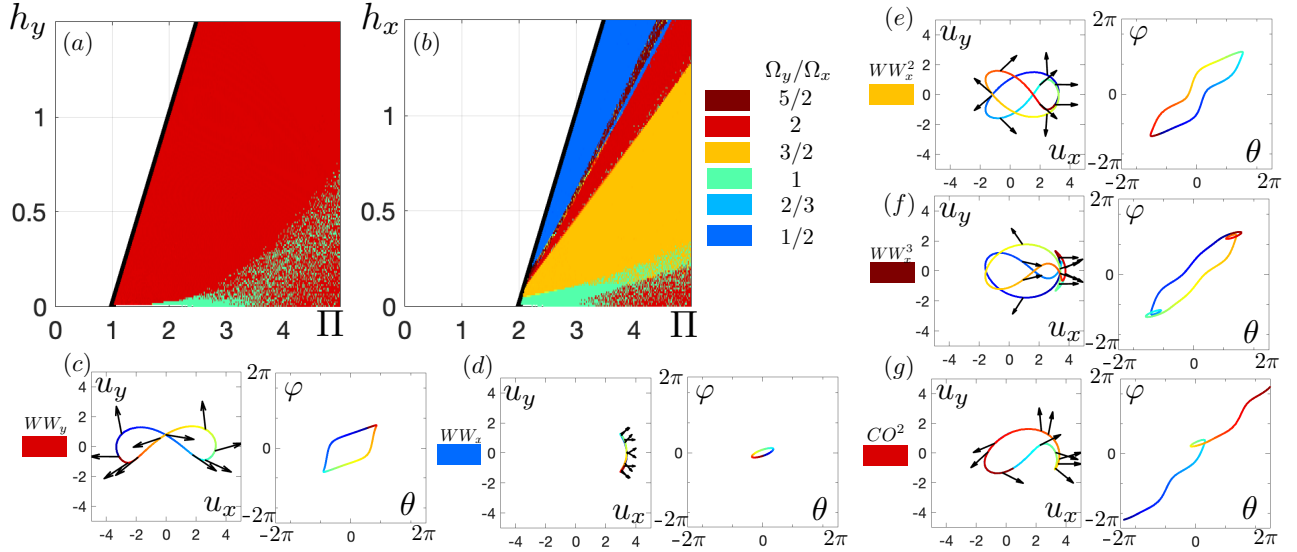

\subsubsection{Adding an external field \texorpdfstring{$\bm{h}$}{h}}

In the non-degenerate case, the orientation of the field matters. We shall discuss the two extreme cases, when the field points in the stiff ($\boldsymbol{h}=h \boldsymbol{\hat{e}}_y$) or the soft ($\boldsymbol{h}=h \boldsymbol{\hat{e}}_x$) direction (Fig.~\ref{fig:1part_nondege_wfield}).
In the light of the variety of solutions already observed in the absence of external field, we don't aim at exhaustively studying the influence of the field on all of them. We concentrate on moderate value of $\Pi/\omega_x^2<5$ and one eccentricity ratio $\omega_y^2 = 2 \omega_x^2$, for which the CO regime in the absence of an external field is a simple elliptic chiral oscillation.

As for the degenerate case, any amount of external field reduces the number of fixed points to two.
The one with the active force pointing in the direction opposite to the field is always linearly unstable.  

$\bullet$ \underline{$\boldsymbol{h}=h_y \boldsymbol{\hat{e}}_y$}. 
When the field points along the stiff direction of the potential, the fixed point polarized in the direction of the field destabilizes linearly for $\Pi>\Pi_c = \omega_x^2+h_y$ via a Hopf bifurcation (Fig. \ref{fig:1part_nondege_wfield}-a).
Both the analysis conducted in the degenerate case and numerical simulations suggest that this transition is supercritical.
The dynamical regime observed for $\Pi\gtrsim \Pi_c$ is extremely similar to that observed in the degenerate case (see Fig.~\ref{fig:1part_dege_field}-b).
Conversely, the non-degeneracy considerably modifies the phase diagram: the domain of existence of the CO regime shrinks and is limited to small fields when $\Pi\gtrsim \omega_x^2$.
For larger $\Pi$ and larger fields, it coexists with a WW$_y$ regime that already exists in the zero-field case (Fig.~\ref{fig:1part_nondege_wfield}-c and~\cite{Damascena2022}).
Numerical inspections indicate that this WW$_y$ regime and the one reported for $\Pi \gtrsim \Pi_c$ connect continuously in a smooth crossover.

 $\bullet$ \underline{$\boldsymbol{h} = h_x \boldsymbol{\hat{e}}_x$}. When the field aligns with the soft direction of the potential, the fixed point polarized in the direction of the field destabilizes linearly for $\Pi>\Pi_c = \omega_y^2 + h_x$, via a Hopf bifurcation (Fig. \ref{fig:1part_nondege_wfield}-b). Here also, the transition is supercritical. The dynamical regime observed for $\Pi\gtrsim \Pi_c$ is a WW regime around the direction of the field, which is now the soft one: this is the WW$_x$ reported in Fig.~\ref{fig:1part_nondege_wfield}-d. Again, the CO regime shrinks and is limited to small fields when $\Pi\gtrsim \omega_y^2$. For larger $\Pi$, yet at small fields, one recovers the WW$_y$ regime reported above (Fig.~\ref{fig:1part_nondege_wfield}-c). However, in contrast with the previous case, the WW regimes observed at low field and large $\Pi$ and the one observed at large field close to $\Pi_c$ are orthogonal. As a result, the smooth crossover taking place in the case of a field aligned with the stiff direction is replaced by a succession of complex dynamics illustrated in Figs.~\ref{fig:1part_nondege_wfield}-e to g.


\section{Coarse-grained dynamics}
\label{sec:cg}

One central observation of our companion paper~\cite{baconnier2025reentrant} is that the transition to the regimes of collective actuation is marked by a reentrance, which is absent from the single particle dynamics described in the previous section.
More specifically, intermediate polarizing fields promote collective actuation, which takes place at lower values of $\Pi$ than in the absence of a field.
At large fields, one recovers the delaying of the transition by the external field observed for the single-particle dynamics.
The reentrant transition can easily be understood from the linear stability analysis of the Frozen Polarized (FP) phase in the framework of the coarse-grained model introduced in~\cite{baconnier2022}.
With an external field, this model generalizes to:
\begin{subequations} 
\label{eq:cg_general}
\begin{align}
\partial_t{\boldsymbol{U}} &= \Pi \boldsymbol{m} + \boldsymbol{F}_{e}[\boldsymbol{U}], \label{eq1:cg} \\
\partial_t{\boldsymbol{m}} &= \left(\boldsymbol{m}\!\times \!\left[ \partial_t{\boldsymbol{U}}+\boldsymbol{h}\right] \right)\!\times \!\boldsymbol{m}\!+\! \frac{1\!-\!\boldsymbol{m}^2}{2}\left(\partial_t{\boldsymbol{U}}+\boldsymbol{h}\right) \! -\!D \boldsymbol{m},
\end{align}
\end{subequations}
where $\boldsymbol{U}(\boldsymbol{r},t)$ and $\boldsymbol{m}(\boldsymbol{r},t)$ are now continuous fields obtained from a local average procedure. The elastic force $\boldsymbol{F}_e \left[\boldsymbol{U}\right]$ is given by the choice of an elastic constitutive relation and the relaxation term $-D \boldsymbol{m}$, with $D>0$, is inherited  from the microscopic angular noise. 

Let us recast here the main steps of the derivation of these equations. The dynamics of the translational degrees of freedom being linear, obtaining its coarse-grained form is straightforward. We then take advantage of the fact that the elastic force being a second-order derivative of the displacement field, it smoothes the displacement field on lengthscales smaller than $l^* \propto \Pi^{-1/2}$, as obtained from the balance of the active driving and the elastic forces.  This considerably simplifies the coarse-graining of the polarity dynamics, as we can safely ignore the fluctuations of the displacement field. Rewriting the deterministic part of the dynamics for the microscopic polarity Eq.~(\ref{eq2:Npart}), using the projector to the normal of $\boldsymbol{\hat{n}}$ and ignoring the fluctuations of the displacement field, one finds:
\begin{equation}
 \partial_{t} \boldsymbol{m} = ( \mathbb{I} - \langle \boldsymbol{\hat{n}}_i \otimes \boldsymbol{\hat{n}}_i \rangle ) (\partial_{t} \boldsymbol{U} + \boldsymbol{h}).
\end{equation}
From invariance by rotation, 
\begin{equation} \label{closure}
 \langle \boldsymbol{\hat{n}}_i \otimes \boldsymbol{\hat{n}}_i \rangle = \varphi(m) \mathbb{I} + \psi(m) \boldsymbol{m} \otimes \boldsymbol{m},
\end{equation}
where $\varphi(m)$ and $\psi(m)$ are two functions of $m$, which must satisfy one additional constraint: since $\Tr(\boldsymbol{\hat{n}}_i \otimes \boldsymbol{\hat{n}}_i ) = 1$, one must have $\Tr \langle \boldsymbol{\hat{n}}_i \otimes \boldsymbol{\hat{n}}_i \rangle = 1$, for any distribution of orientations.
When $m=0$, this constraint imposes $\varphi(0)=1/2$. When $m=1$, the equality of all $ \boldsymbol{\hat{n}}_i $ imposes $\psi(1) = 1$ and $\varphi(1) = 0$.
As a simple ansatz, we write $\langle \boldsymbol{\hat{n}}_i \otimes \boldsymbol{\hat{n}}_i \rangle$ as the only second-order polynomial in $m$ that is compatible with the above constraints:
\begin{equation}
 \langle \boldsymbol{\hat{n}}_i \otimes \boldsymbol{\hat{n}}_i \rangle = \frac{1 - m^{2}}{2}\mathbb{I} + \boldsymbol{m} \otimes \boldsymbol{m}.
\end{equation}
The rotational noise acting on the $\boldsymbol{\hat{n}}_i$'s simply coarse-grains into $-D \boldsymbol{m}$, following Itô calculus or the approach introduced in~\cite{Lam2015a}. A few lines of calculations then lead to Eqs.~(\ref{eq:cg_general}).

In principle, these equations must be completed by boundary conditions where the displacements or forces are specified.
However, we shall consider here a simpler mean-field version of the model, assuming that the dynamics condensates on two spatially homogeneous modes of stiffnesses $\omega_x^2$ and $\omega_y^2$, as observed both experimentally and numerically whenever collective actuation sets in~\cite{baconnier2022, Gu2025}. Note that, for $\omega_{x}^{2} = \omega_{y}^{2}$, this convention virtually represents a two-dimensional elastic sheet with adhesion to the substrate \cite{Xu2023}. The above equations then simplify into:
\begin{subequations} 
\label{eq:cg_2modes_general}
\begin{align}
\dot U_x &= \Pi\,m_x - \omega_x^2 U_x, \label{eq3:cgx} \\
\dot U_y &= \Pi \,m_y - \omega_y^2 U_y, \label{eq3:cgy} \\
\dot m_x &= \frac{1- m_x^2 + m_y^2}{2}\left(\dot U_x +h_x\right) \nonumber\\ &\qquad - m_x m_y \left(\dot U_y +h_y\right)-D m_x, \label{eq3:cgmx} \\
\dot m_y &= -m_x m_y \left(\dot U_x +h_x\right) \nonumber\\ &\qquad + \frac{1+ m_x^2 - m_y^2}{2} \left(\dot U_y+h_y\right)-D m_y. \label{eq3:cgmy}
\end{align}
\end{subequations}

In the following, we shall investigate the phase diagram for the above dynamics in the degenerate $(\omega^2_x = \omega^2_y)$ and non-degenerate $(\omega^2_x < \omega^2_y)$ cases, in the presence and absence of external field, and considering $D>0$.

\subsection{The degenerate case: $\omega^2_x = \omega^2_y = \omega^2_0$}

\subsubsection{In the absence of an external field \texorpdfstring{$\bm{h}=\bm{0}$}{h=0}}

In contrast with the single-particle description, here, due to the presence of the relaxation term $-D \boldsymbol{m}$, the only fixed point is $(\boldsymbol{U}=\boldsymbol{0}, \boldsymbol{m}=\boldsymbol{0})$, which describes a disordered phase.
This fixed point is linearly stable for $\Pi<\Pi_c = 2(\omega_0^2+D)$.
\begin{figure}[t!]
\centering

\begin{tikzpicture}

\node[rotate=0] at (0.0,0.0) {\includegraphics[width=0.8\columnwidth]{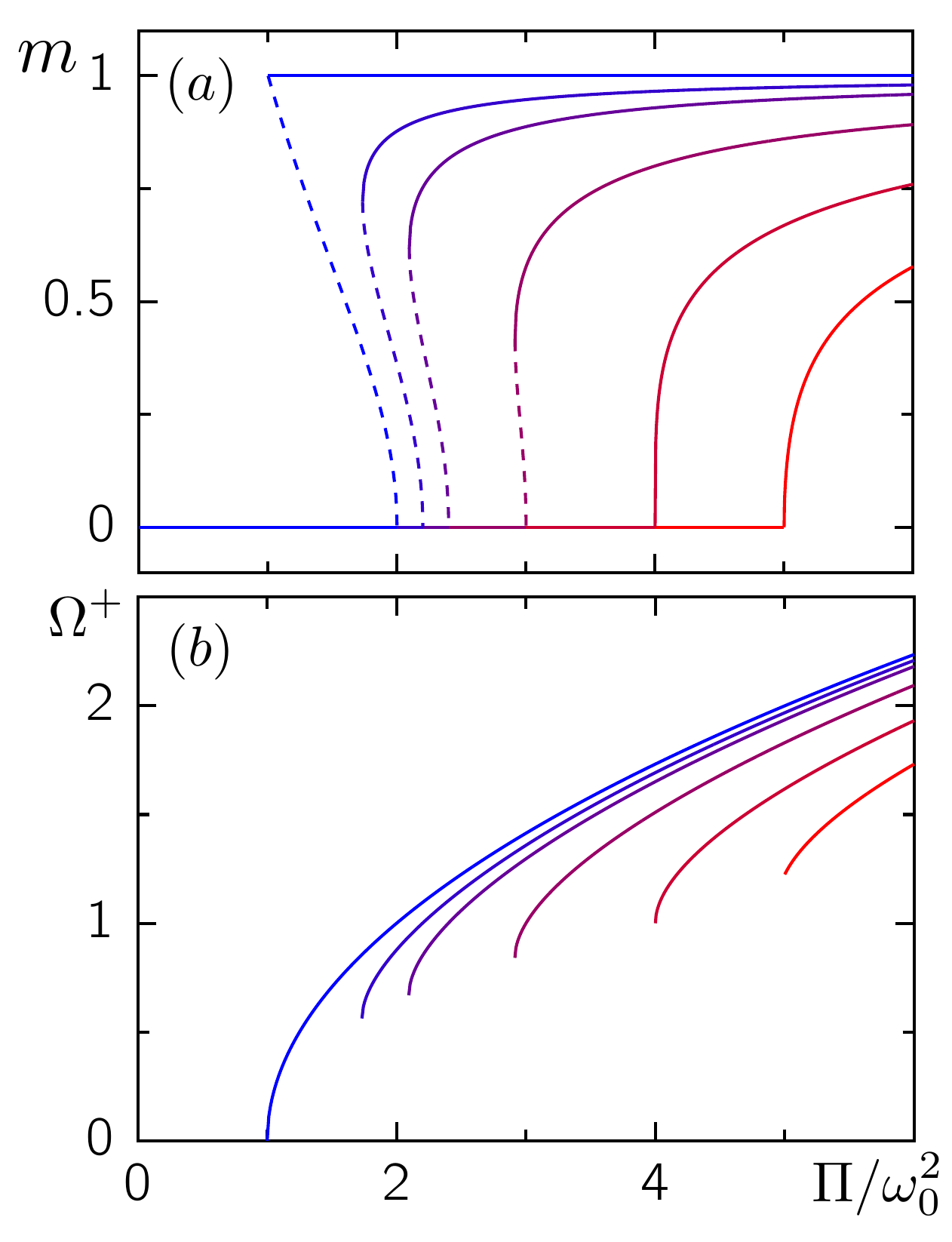}};

\draw[->] (2.1,4.1) -- (2.62,1.35);
\node[] at (2.72,1.1) {\small $D$};

\end{tikzpicture}
\vspace{-0.5cm}
\caption{{\bf Coarse-grained dynamics in the degenerate case and the absence of an external field:} (a) Amplitude of the polarization $m$ as a function of $\Pi/\omega_0^2$, as given by Eq. (\ref{eq:cg_dege_h0_rot_mag}); the solid, respectively dashed line indicates a linearly stable ($m^+$), respectively unstable ($m^-$) solution. (b) Rotation rate of the linearly stable solution $\Omega^+$ for the same values of $\Pi/\omega_0^2$, as given by Eq. (\ref{eq:cg_dege_h0_rot_omega}). The colors indicate the amplitude of the relaxation term: from blue to red $D = \{0, 0.1, 0.2, 0.5, 1, 1.5\}$; $\omega_{0}^{2} = 1$.}
\label{fig:CG_dege_nofield}
\end{figure}
At threshold, a Hopf bifurcation takes place. From invariance by rotation, one can, as for the single particle, rewrite the dynamics in polar coordinates $\boldsymbol{U}=(R\cos\varphi,R\sin\varphi)$ and $\boldsymbol{m}=(m\cos\theta,m\sin\theta)$:
\begin{subequations}
\label{eq:cg_nofield_degenrated}
\begin{align}
\dot{R} &= \Pi\, m \cos\gamma - \omega_0^2 R, \\
\dot{m} &= \frac{1-m^2}{2}\left(\Pi\,m - \omega_0^2 R \cos\gamma \right) -Dm, \\
\dot{\gamma} &= \left(\frac{1+m^2}{2m} \omega_0^2 R - \frac{\Pi\, m}{R} \right) \sin\gamma, \\
\dot{\theta} &= \frac{1+m^2}{2m} \omega_0^2 R \sin\gamma,
\end{align}
\end{subequations}
with $\gamma = \theta - \varphi$. This allows for an exact computation of the steady rotating solutions. One finds that two orbiting solutions emerge from a saddle node bifurcation, taking place at $\Pi_\textrm{sn} = \left(\omega_0 + \sqrt{D}\right)^2$ (see Fig.~\ref{fig:CG_dege_nofield}):
\begin{subequations}
\label{eq:cg_dege_h0_rot}
\begin{align}
    m_{\pm} &= \sqrt{\frac{\omega_0^2 - D}{\Pi} \pm \sqrt{\left(\frac{\omega_0^2 - D}{\Pi}\right)^2 - 2\frac{\omega_0^2 + D}{\Pi} +1 }}, 
    \label{eq:cg_dege_h0_rot_mag}\\
    \Omega_{\pm} &= \omega_0^2 \sqrt{\frac{\Pi (1+m_{\pm}^2)}{2\omega_0^2} - 1}. \label{eq:cg_dege_h0_rot_omega}
\end{align}    
\end{subequations}
The $m_{+}$ solution is linearly stable, while $m_{-}$ is unstable.
The polarization at the saddle node is $m_\textrm{sn}=\sqrt{\frac{\omega_0 - \sqrt{D}}{\omega_0 + \sqrt{D}}}$. We thus find that the nature of the bifurcation from the disordered state depends on $D$. When $D<\omega_0^2$, the transition is subcritical and the polarization jumps discontinuously from zero in the disordered state to a finite value in the rotating phase.  Conversely, when $D\ge\omega_0^2$, the stable orbiting solution branches off continuously from the zero-polarization disordered state.
Surprisingly, except when $D\rightarrow0$, the rotation rate is always finite at the transition. When the transition is discontinuous, the rotation rate at the saddle node is $\Omega_\textrm{sn}=\omega_0^2 ( D/\omega_0^2 )^{1/4}$. When it is continuous, the rotation rate at the transition is $\Omega_{c}=\omega_0^2 ( D/\omega_0^2 )^{1/2}$. Note that this transition scenario is in sharp contarst with the noiseless single-particle, where the transition to CO is supercritical.

It is interesting to elaborate on the connections between the zero-field coarse-grained equations and their chiral oscillating solutions \cite{baconnier2022}, and recent theoretical work on modeling of dense pedestrian crowds \cite{Gu2025}. Indeed, the mean-field equations proposed in \cite{Gu2025} from symmetry considerations for the dynamics of the displacement and polarity of the crowd are formally equivalent to Eqs. (\ref{eq:cg_general}) at the linear level. The difference between the two models comes from the nonlinear terms, which, in Eqs. (\ref{eq:cg_general}), are obtained from the local averaging procedure. First, Gu \textit{et al.} considered that the coefficient in front of the nonlinear reorientation term $(\boldsymbol{m} \times \partial_t \boldsymbol{U}) \times \boldsymbol{m}$ is negative. Moreover, in Eqs. (\ref{eq:cg_general}), the term in front of the $\partial_t \boldsymbol{U}$ term exhibits a nonlinear saturation $(1 - \boldsymbol{m}^2)/2$. The emergence of chiral oscillations in dense crowds \cite{Gu2025} and in polar active solids \cite{baconnier2022} therefore seem to share the same mechanism.

\subsubsection{Adding an external field \texorpdfstring{$\bm{h}$}{h}}

\begin{figure}[b!]
\centering
\begin{tikzpicture}

\node[rotate=0] at (0.0,0.0) {\includegraphics[width=0.9\columnwidth]{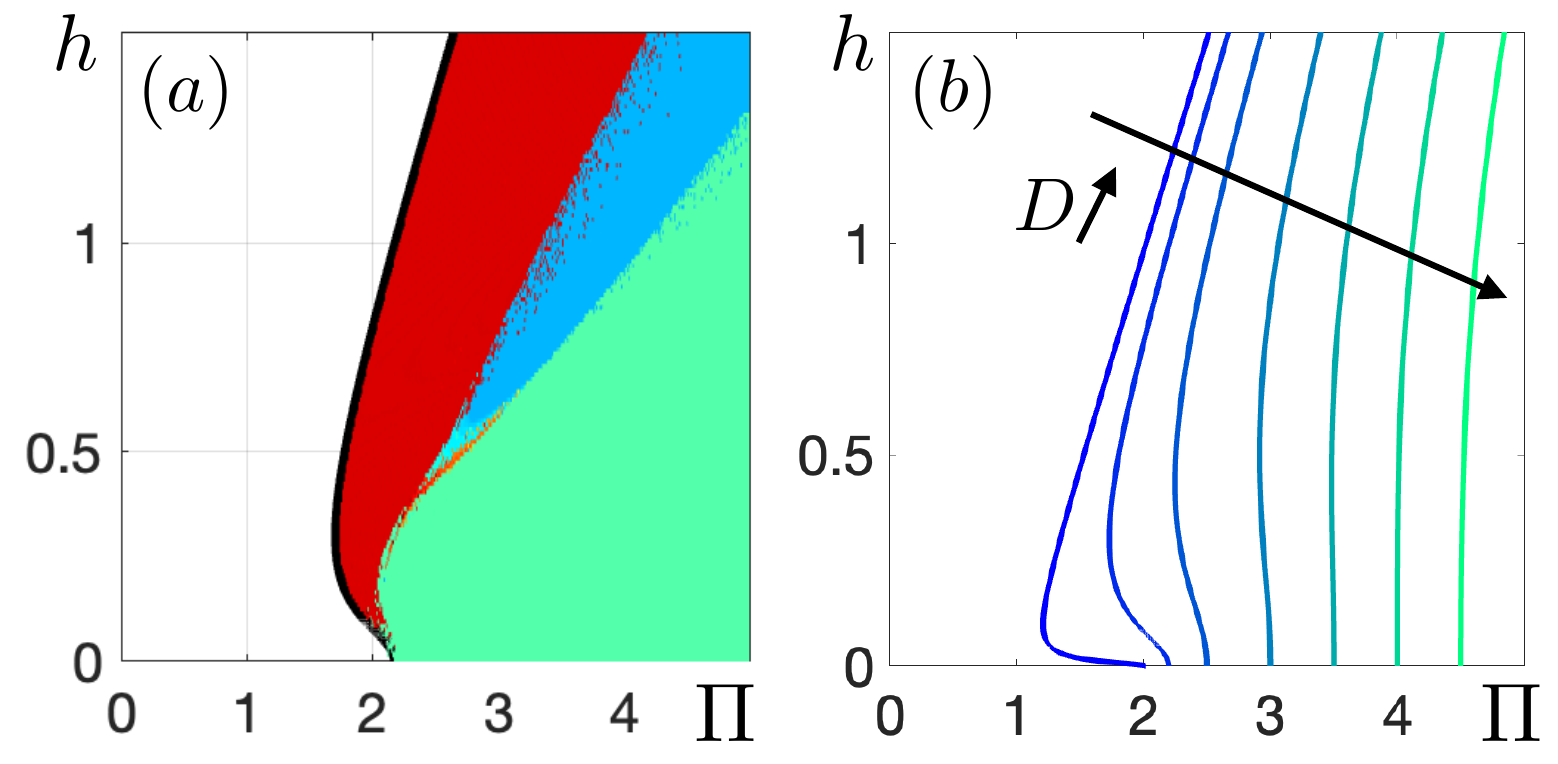}};

\node[rotate=0] at (-0.49,-1.19) {\small \textbf{CO}};

\node[rotate=73] at (-1.94,1.4) {\small \textbf{FP}};
\node[rotate=73] at (-1.52,1.22) {\small \textbf{WW}$_y$};

\node[rotate=57] at (-0.61,1.2) {\small \textbf{WW}$^2_y$};

\end{tikzpicture}
\vspace{-0.3cm}
\caption{{\bf Coarse-grained dynamics in a degenerate potential, $(\omega_x^2=\omega_y^2=1)$ with a field:} (a) Phase diagram for $D=0.1$; the color code and the type of dynamics is the same as for the single particle (see Fig.~\ref{fig:1part_dege_field}a). The thick black line indicates the linear stability threshold for the fixed point corresponding to the FP state. (b) Stability threshold of this fixed point for $D=\{0.01, 0.1, 0.25, 0.5, 0.75, 1, 1.25\}$.}
\label{fig:CG_dege_wfield}
\end{figure}
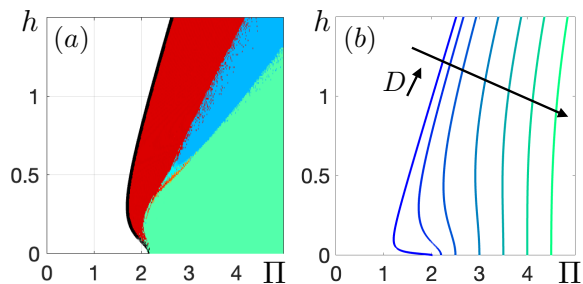

Figure~\ref{fig:CG_dege_wfield}-a displays the phase diagram obtained numerically, solving Eqs.~(\ref{eq:cg_general}), for $\omega_0^2=1$, $\hb = h\hat{\bf e}_y$ and $D=0.1$.
One recovers the same four dynamics reported in the case of a single particle. For small enough $\Pi$, the system is frozen and polarized in the direction of the field.
This state is described by the unique stable fixed point when $h>0$, which reads 
$(U_x = 0, U_y = \Pi m_y / \omega_0^2, m_x = 0, m_y = \sqrt{1+(D/h)^2}-D/h)$.
The Jacobian, evaluated at the fixed point, diagonalizes by blocks in the $x$ and $y$ spaces. The fixed point destabilizes via a Hopf bifurcation that is controlled by the eigenvalue associated with the $y$ subspace. The resulting threshold reads:
\begin{equation}\label{eq:Pic_cg_dege_field}
\Pi_c(h,D) =\frac{h^2 \left(\omega_0^2 + \sqrt{D^2+h^2}\right)}{D^2+h^2-D \sqrt{D^2+h^2}},
\end{equation}
and is illustrated for increasing values of $D$ in Fig.~\ref{fig:CG_dege_wfield}-b.

The central observation to be made is that the phase diagram for the coarse-grained dynamics at low enough noise $D<\omega_0^2$ exhibits a reentrance transition from the frozen to the oscillating phases.
Intutitively this can be understood as follows: for low enough field, the polarization of the frozen phase ``helps'' the onset of the collective actuation by introducing some level of ordering in the orientation of the active forces; conversely, for large field the polarization opposes the collective actuation by imposing strongly one specific orientation to the active forces. 

There is a correspondence between the stability threshold of the FP state in the presence of a field (Eq.~(\ref{eq:Pic_cg_dege_field})) and the rotating solutions in the absence of a field obtained in the previous section (Eqs.~(\ref{eq:cg_dege_h0_rot})).
Indeed, using the polarization of the FP state $m$ instead of the polarizing field $h$ in the expression~(\ref{eq:Pic_cg_dege_field}) for the stability threshold, and then solving for the polarization $m$ in the resulting equation, we recover Eq.~(\ref{eq:cg_dege_h0_rot_mag}).
As a consequence, Fig.~\ref{fig:CG_dege_nofield}-a and Fig.~\ref{fig:CG_dege_wfield}-b are equivalent under the relation between $m$ and $h$ in the FP state.
This means that the activity $\Pi$ necessary to rotate a state with polarization $m$ in the absence of a field is identical to the activity necessary to destabilize a frozen state with a polarization $m$ resulting from an external field.

Once collective actuation sets in, one recovers essentially the same dynamics as for the single particle, namely the Chiral Oscillation (CO) at low enough field, and the Windscreen Wipers (WW$^2_y$ and WW$_y$) at larger fields.
To date, no analytical solution for the boundaries separating the domain of existence of these dynamics has been obtained.

%

\begin{figure*}[t!]
\centering

\begin{tikzpicture}

\node[rotate=0] at (7.55,0.0) {\includegraphics[width=0.9\textwidth]{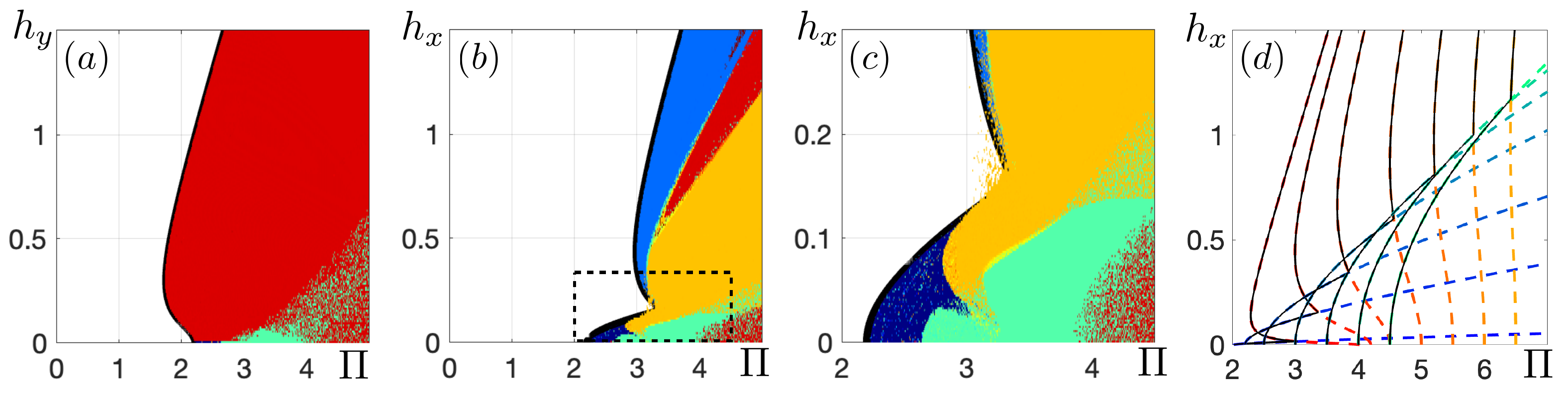}};

\node[] at (1.85,-0.62) {\footnotesize \textbf{NICA}};
\draw[] (1.55,-0.80) -- (2.15,-0.80);
\draw[->] (1.85,-0.80) -- (1.65,-1.42);

\draw[] (1.62,-1.47) ellipse (0.21 and 0.09);

\node[] at (2.60,-0.21) {\footnotesize \textbf{CO}};
\draw[] (2.30,-0.38) -- (2.90,-0.38);
\draw[->] (2.60,-0.38) -- (2.30,-1.40);

\node[rotate=73] at (1.46,1.38) {\small \textbf{FP}};
\node[rotate=73] at (1.88,1.20) {\small \textbf{WW}$_y$};

\node[rotate=75] at (6.18,1.38) {\small \textbf{FP}};
\node[rotate=75] at (6.60,1.20) {\small \textbf{WW}$_x$};

\node[rotate=33] at (10.2,-0.65) {\footnotesize \textbf{CO}};
\node[rotate=0] at (10.82,1.45) {\small \textbf{WW}$^2_x$};

\node[] at (8.7,-0.04) {\footnotesize \textbf{NICA}};
\draw[] (8.4,-0.22) -- (9.0,-0.22);
\draw[->] (8.7,-0.22) -- (8.95,-0.89);

\end{tikzpicture}
\vspace{-0.35cm}
\caption{{\bf Coarse-grained dynamics in a non-degenerate potential, $(\omega_x^2=1, \omega_y^2=2)$ in the presence of an external field:} (a) Phase diagram with the field in the stiff direction $h = h_y{\bf \hat{e}}_y$; the solid black line indicates $\Pi_c(h_y)$ as given by Eq. (\ref{eq:threshold_cg_stiff}). (b) Phase diagram with the field in the soft direction $h = h_x{\bf \hat{e}}_x$; the solid black line indicates $\Pi_c(h_x)$ as given by Eqs. (\ref{eq:threshold_cg_soft}). In (a-b), the color codes for the value of $\Omega_y/\Omega_x$ as in Fig.~\ref{fig:1part_nondege_wfield}. The dark blue domain corresponds to the NICA dynamics, which is confined to the $x$ direction; (c) Zoom on the small fields, close to the destabilization of the fixed point, as indicated by the rectangle dashed lines in panel (b). (d) Limits of linear stability of the Frozen Polarized (FP) state for increasing values of $D=0.01, 0.1, 0.25,0.5, 0.75, 1, 1.25$: the red to orange dashed lines indicate $\Pi_c^{(1)}(h_x,D)$ and the blue to green dashed lines indicate $\Pi_c^{(2)}(h_x,D)$; the black continuous overlaid indicate $\Pi_c(h_x,D)=\min\left(\Pi_c^{(1)},\Pi^{(2)}_c\right)$ (see text for details).}
\label{fig:CG_nondege_wfield}
\end{figure*}

\subsection{The non-degenerate case $\omega^2_x < \omega^2_y$}

\subsubsection{In the absence of an external field \texorpdfstring{$\bm{h}=\bm{0}$}{h=0}}

As for the degenerate case, the relaxation term $-D\boldsymbol{m}$ imposes the uniqueness of the fixed point $({\boldsymbol{U}=\boldsymbol{0},\boldsymbol{m}=\boldsymbol{0}})$ describing the disordered solution. Although the isotropy of the dynamics is broken, the $x$ and $y$ directions still decouple in the linearized dynamics around the fixed point and one easily finds that the fixed point destabilizes, following a Hopf bifurcation, along the softest direction when $\Pi>2(\omega_x^2+D)$.

The oscillating dynamics selected at the Hopf bifurcation is rather specific as the oscillation only takes place along the soft, here the $x$ direction. The frequency of the oscillation $\Omega \sim D^{1/2}$ and a weakly nonlinear analysis demonstrates the supercritical nature of the bifurcation.
This is the Noise-Induced Collective Actuation dynamics discussed thoroughly in Ref.~\cite{baconnier2024noise}.
The NICA regime is replaced by the CO regime for larger values of $\Pi$, yet smaller than  $2(\omega_y^2+D)$. However, the stability of the NICA regime in the coarse-grained dynamics has not been investigated.

\subsubsection{Adding an external field \texorpdfstring{$\bm{h}$}{h}}

Here also, because of the non-degeneracy, the orientation of the field matters and we shall analyze the two extreme cases, when the field points in the stiff ($\boldsymbol{h} = h_y \boldsymbol{\hat{e}}_y$) or the soft ($\boldsymbol{h} = h_x \boldsymbol{\hat{e}}_x$) direction.
We again concentrate on moderate values of $\Pi/\omega_x^2<5$ and one eccentricity ratio $\omega_y^2 = 2 \omega_x^2$, for which the CO regime in the absence of an external field is a simple elliptic chiral oscillation.

As for the degenerate case, at small enough $\Pi$, there is only one linearly stable fixed point, corresponding to the FP state in the direction of the imposed field, and the Jacobian, evaluated at the fixed point, diagonalizes by blocks in the $x$ and $y$ space.

$\bullet$ \underline{$ \boldsymbol{h} = h_y \boldsymbol{\hat{e}}_y$}.  The fixed point $(U_x\!=\!0$, $U_y\!=\!\Pi m_y/\omega_y^2$, $m_x\!=\!0$, $m_y\!=\!\sqrt{1+(D/h_y)^2}-D/h_y)$ destabilizes via a Hopf bifurcation that is controlled by the eigenvalue associated with the $x$ subspace.
From the point of view of the linear instability, the situation is thus perfectly analogous to the degenerate case, with $\omega_x^2$ replacing $\omega_0^2$:
\begin{equation} \label{eq:threshold_cg_stiff}
    \Pi_c(h_y,D) =\frac{h_y^2 \left(\omega_x^2 + \sqrt{D^2+h_y^2}\right)}{D^2+h_y^2 - D \sqrt{D^2+h_y^2}}.
\end{equation}
The phase diagram (Fig.~\ref{fig:CG_nondege_wfield}-a) inherits both the properties of the single particle and the coarse-grained dynamics. On one hand, there is a strong shrinkage of the domain of existence of the CO regime, as compared to the degenerate case, and a smooth crossover across WW$_y$ regimes from small to large fields. On the other hand, one observes the reentrance transition when increasing $h$ for $\Pi \lesssim \Pi_c(h=0)$.

 $\bullet$ \underline{$\boldsymbol{h} = h_x \boldsymbol{\hat{e}}_x$}. The fixed point polarized in the direction of the field $(U_x\!=\!\Pi m_x/\omega_x^2$,$U_y\!=\! 0$,$m_x\!=\!\sqrt{1+(D/h_x)^2}-D/h_x$,$m_y\!=\!0)$ destabilizes via a Hopf bifurcation. However, in contrast with all previous cases, the real part of both eigenvalues cross each other when varying $\Pi$ or $h$. As illustrated in Fig.~\ref{fig:CG_nondege_wfield}-d for increasing values of $D$, we now have $\Pi_c(h_x,D)=\min\left(\Pi_c^{(1)},\Pi^{(2)}_c\right)$, with:
\begin{subequations} \label{eq:threshold_cg_soft}
\begin{align}
\Pi_c^{(1)}(h_x,D) & =  \frac{h_x^2 \left(\omega_y^2 + \sqrt{D^2+h_x^2}\right)}{D^2+h_x^2 - D \sqrt{D^2+h_x^2}}, \\ 
\Pi_c^{(2)}(h_x,D) & =  \frac{h_x^2 \left(\omega_x^2 + \sqrt{D^2+h_x^2}\right)}{D \sqrt{D^2+h_x^2}-D^2},
\end{align}
\end{subequations}
resulting in a far more complex phase diagram (Fig.~\ref{fig:CG_nondege_wfield}-b). For low enough noise, the intersection of the real parts of the eigenvalues leads to the presence of a cusp in the stability boundary, leading eventually to a double reentrance. These reentrance not only concern the transition from the FP state to the collective actuation ones, they also shape the complex transition and coexistence lines between the various oscillating regimes (Fig.~\ref{fig:CG_nondege_wfield}-c). We also note the dark blue region in the phase diagram, illustrating the extension of the NICA regime to finite fields, a feature that is absent when the field acts in the stiff direction.

\section{Conclusion}

An external field acting on the orientation of the active forces in a polar active solid gives rise to a rich diversity of oscillating dynamics.
In this work, we have explored and described them at the single-particle and coarse-grained levels. In both cases, when the modes selected by the active dynamics are degenerate, analytical results are available, even in the presence of a field. However, when the modes are non-degenerate, the complexity of the oscillating regimes prevents analytical descriptions, and one must rely on numerical simulations to characterize the regimes and the transitions between them.

Much more work remains to be done to understand thoroughly the non-degenerate case in the presence of an external field, especially the oscillating regimes emerging for large stiffness ratio between the selected modes \cite{Damascena2022}.
The transition to collective actuation in large systems in the presence of an external field is another matter of interest.
In Ref.~\cite{baconnier2022}, it was shown that the phase space coexistence of disordered and CO dynamics at the coarse-grained level translates, in finite systems, into their spatial coexistence during the transition to collective actuation. Here, the recurrent coexistence of different oscillating regimes begs the question of the possibility of their spatial coexistence in a large polar active solid.

The structure of Eqs. (\ref{eq:Npart_general}) and (\ref{eq:cg_general}) reveal that an external polarizing field $\boldsymbol{h}$ is equivalent to a change of Galilean frame of reference moving at constant velocity $\boldsymbol{V} \propto \boldsymbol{h}$. This connection between polarizing fields and mechanical driving thus raises the question of the response of active solids to applied mechanical stresses, and of the possible emergence of odd elastic moduli in such systems, which is ripe for further research.

Finally, if such polarizing external fields could be engineered or evidenced in biological contexts, such as in confined cell monolayers \cite{Peyret2019, petrolli2019confinement}, or dense bacterial suspensions \cite{chen2017weak, liu2021viscoelastic} and bio-films \cite{Xu2023}, our work could help to uncover new mechanisms for oscillatory dynamics and regulation in living systems.

\begin{acknowledgments}
We acknowledge multiple discussions with our colleagues, Michel Fruchart, Gustavo D\"uring, Etienne Fodor, and new students Paul Bernard, Benjamin Buhl, and Qingju Qi, whose curiosity and motivation drove the effort to summarize the above results in one self-consistent paper. We acknowledge financial support from Ecole Doctorale ED564 Physique en Ile de France for PB's Ph.D. grant.
\end{acknowledgments}

\bibliographystyle{apsrev4-1} 

\bibliography{sapam.bib}

\appendix

\section{Single particle in the degenerate case: Mapping of the dynamics on the weighting pendulum, close to the exceptional point}

\label{app:pendulum}

Starting from Eqs. (\ref{eq:1part_degenerated}), we define $\varepsilon = (\Pi - \omega_{0}^{2})/\omega_{0}^{2}$, $\rho = R - 1$, $H = h/\omega_{0}^{2}$. Rescaling time by $\omega_{0}^{2}$, we obtain:
\begin{subequations}
\begin{align}
 \dot{\rho}  &= \left( 1 + \varepsilon \right) \cos(\gamma) - 1 - \rho, \\
\dot{\varphi}  &= \frac{1 + \varepsilon}{1 + \rho} \sin(\gamma),  \\
\dot{\gamma} &= \left( \rho - \frac{\rho - \varepsilon}{\rho + 1} \right) \sin(\gamma) - H \sin(\gamma + \varphi).
\end{align}
\end{subequations}
\noindent The scaling of the transition lines $\Pi_c$ and $\Pi^{\star}$ suggests different limit behavior when $\varepsilon \rightarrow 0$ with $\tilde{H} = H/\varepsilon$ constant: FP for $\tilde{H} > 1$, WW$_y$ for $1/3 < \tilde{H} < 1$, and CO for $\tilde{H} < 1/3$. To identify the asymptotic solutions to the equations of motion, we rescale the different quantities with $\varepsilon$. The case without external field suggests:
\begin{subequations}
\begin{align}
 \rho(t)  &= \varepsilon \tilde{\rho} (\sqrt{\varepsilon}t), \\
 \varphi(t)  &= \tilde{\varphi} (\sqrt{\varepsilon}t),  \\
 \gamma(t)  &= \sqrt{\varepsilon} \tilde{\gamma} (\sqrt{\varepsilon}t).
\end{align}
\end{subequations}
Inserting these scaling forms in the equations of motion, we next separate the different orders in $\varepsilon$ in the limit $\varepsilon \rightarrow 0$. At zeroth order, we find:
\begin{subequations} \label{eq:zeroth_order_pendulum_app}
\begin{align}
 \tilde{\rho}  &= 1 - \frac{\tilde{\gamma}^2}{2}, \label{eq1:zeroth_order_pendulum_app} \\
\dot{\tilde{\varphi}}  &= \tilde{\gamma}, \label{eq2:zeroth_order_pendulum_app} \\
\dot{\tilde{\gamma}} &= - \tilde{H} \sin \tilde{\varphi}. \label{eq3:zeroth_order_pendulum_app}
\end{align}
\end{subequations}
$\tilde{\rho}$ is actually a fast variable, which relaxes instantaneously to $1 - \tilde{\gamma}^2 / 2$ in the limit $\varepsilon \rightarrow 0$ (Eq. (\ref{eq1:zeroth_order_pendulum_app})). The equations for $\tilde{\varphi}$ and $\tilde{\gamma}$ describe a pendulum (Eqs. (\ref{eq2:zeroth_order_pendulum_app}) and (\ref{eq3:zeroth_order_pendulum_app})): $\ddot{\tilde{\varphi}} = - \tilde{H} \sin \tilde{\varphi}$.
At this order, the energy
\begin{equation}
 E = \frac{\tilde{\gamma}^{2}}{2} - \tilde{H} \cos \tilde{\varphi},
\end{equation}
is conserved and there is no mechanism to select the orbit.

At the next order in $\varepsilon$, Eq.~(\ref{eq3:zeroth_order_pendulum_app}) becomes
\begin{equation}
  \dot{\tilde{\gamma}} = - \tilde{H} \sin \tilde{\varphi}+ \sqrt{\varepsilon} \tilde{\gamma} \left(1- \tilde{\gamma}^2 - \tilde{H} \cos \tilde\varphi \right). 
\end{equation}
This new term generates an energy drift:
\begin{subequations}
  \label{eq:general_energy_drift_app}
 \begin{align}
 \dot{E} &= \sqrt{\varepsilon} \tilde{\gamma}^{2} \left[ 1 - \tilde{\gamma}^2 - \tilde{H} \cos \tilde{\varphi} \right], \\
 &= 2 \sqrt{\varepsilon} \left[ E + \tilde{H} \cos \tilde{\varphi} \right] \left[ 1 - 2E -3\tilde{H} \cos \tilde{\varphi} \right].
 \end{align}
\end{subequations}
%
Close enough to the exceptional point, the drift is slow, allowing to define the energy change over one period:
\begin{equation} \label{eq:_energy_drift_averaged_app}
 \delta E = \int_0^T \dot{E}(t)dt.
\end{equation}
For a given value of $\tilde{H}$, to any orbit of energy $E$ is associated an averaged energy drift $\delta E(\tilde{H}, E)$. 
If, for such orbit, $\delta E > 0$ (resp. $\delta E < 0$), the energy drift increases (resp. decreases) energy over time.
Equilibrium orbits satisfy $\delta E = 0$, and stable ones require $\partial \delta E/\partial E < 0$.

We can compute $\delta E$ in the limiting cases of very large and very small energies, and for the heteroclinic orbit $E = \tilde{H}$.
When $E \gg \tilde{H}$ and $E\gg 1$, we are in the fast chiral state with $\tilde{\varphi}(t) \simeq \sqrt{2E}t$, and the energy change over a period is, at leading order, $\delta E=-4\pi\sqrt{2\epsilon}E^{3/2}<0$. 
On the contrary, when $E+\tilde H \ll \tilde{H}$, the amplitude of the oscillations is small and we can expand the cosine in Eq.~(\ref{eq:general_energy_drift_app}).
At leading order, we get: $\delta E = 2\pi\sqrt{\varepsilon/\tilde H} (1 - \tilde{H})(E+\tilde{H})$.
This result shows that the state $E=-\tilde H$ is stable for $\tilde H>1$ and unstable for $\tilde H<1$; however, it does not provide the amplitude of the oscillations close to the threshold.
These can be obtained by expanding at the next order:
taking $\tilde \varphi(t)=A\cos(\omega t)$, with $\omega=\sqrt{\tilde H}$, we obtain for the energy drift:
\begin{equation}
    \delta E = \sqrt{\varepsilon}\tilde H A^2 T\left[\frac{1}{2}(1-\tilde H)-\frac{5}{16}\tilde H A^2\right].
\end{equation}
The amplitude $A$ is obtained by solving $\delta E=0$, leading to 
\begin{equation}
    A = \sqrt{\frac{8}{5}\frac{1-\tilde H}{\tilde H}}
    = \sqrt{\frac{8}{5}\frac{\Pi-\omega_0^2 -h}{h}}.
\end{equation}
For the frequency, we note that $\varphi(t)=\tilde\varphi\left(\sqrt{\varepsilon} t\right)
=A\cos\left(\sqrt{H} t\right)$.
Altogether, we obtain in the small amplitude regime:
\begin{equation}\label{eq:exp_exceptional_comp}
    \varphi(t)\simeq \sqrt{\frac{8}{5}\frac{\Pi-\omega_0^2 -h}{h}}
    \cos\left(\sqrt{\frac{h}{\omega_0^2}} t\right).
\end{equation}

On the heteroclinic orbit, $E=\tilde H$, the energy change is:

\begin{align}
  \delta E &= \sqrt{2\varepsilon \tilde H} \int_{0}^{2\pi} \sqrt{1+\cos \tilde{\varphi}} \left( 1 - 2\tilde{H} - 3\tilde{H} \cos \tilde{\varphi} \right) d\varphi,\nonumber\\
  & = 8 \sqrt{\epsilon\tilde{H}} (1 - 3\tilde{H}).
\end{align}
Hence, if $\tilde{H} > 1/3$, the energy decays on the heteroclinic orbit; this is the WW$_y$ regime.
In contrast, if $\tilde{H} < 1/3$, the energy increases on the heteroclinic orbit: this is the CO regime.

Finally, we can compute the energy change $\delta E(\tilde{H},E)$ numerically, using Eq.~(\ref{eq:general_energy_drift_app}) and the exact expressions of the pendulum solutions~\cite{belendez2007exact} (Fig. \ref{fig:mapping_pendulum}-a).

\section{Single particle in the degenerate case: Weakly nonlinear WW regime of the single particle}
\label{app:multiple_scale_ww}

We start from the equations for $R$, $\varphi$, and $\theta=\gamma+\varphi$ (Eqs. (\ref{eq:1part_degenerated})), where we set $\omega_0^2=1$:
\begin{subequations}
\begin{align}
\dot{R} &= \Pi \cos (\theta - \varphi) - R, \\
\dot{\varphi} &= \frac{\Pi}{R} \sin (\theta - \varphi), \\
\dot{\theta} &= R \sin (\theta - \varphi) - h \sin \theta.
\end{align}
\end{subequations}

The linear stability analysis of the state ($R = \Pi$, $\theta = \varphi = 0$) gives the instability threshold $\Pi_c = 1 + h$.
To study the weakly nonlinear regime close to the instability, we take $\Pi = \Pi_c + \varepsilon$, where $\varepsilon \ll 1$.
We introduce the slow timescale $T = \varepsilon t$, so that $\partial_t$ becomes $\partial_t + \varepsilon \partial_T$, and the expansions:
\begin{subequations}
\begin{align}
R (t) &= \sum_{k \geq 0} \varepsilon^k R_k(t, T), \label{eq:WW_weakly_nonlinear_expansion_R} \\
\varphi(t) &= \varepsilon^{1/2} \sum_{k \geq 0} \varepsilon^k \varphi_k(t, T), \label{eq:WW_weakly_nonlinear_expansion_phi} \\
\theta(t) &= \varepsilon^{1/2} \sum_{k \geq 0} \varepsilon^k \theta_k(t, T) \label{eq:WW_weakly_nonlinear_expansion_theta}.
\end{align}
\end{subequations}

The exponents are suggested by numerical simulations, but they can also be determined from the equations.

\subsection{Orders $0$ and $1/2$}

At order $0$, only the equation on $R$ is nontrivial and it reduces to $\partial_t R_0 = \Pi_c - R_0$, leading to the constant solution $R_0 = \Pi_c$ after a time $1$.
The lowest order for the dynamics of $\varphi$ and $\theta$ is the order $1/2$. Using $R_0 = \Pi_c = 1 + h$ in these equations, we get:
%
\begin{equation}
 \partial_t \begin{pmatrix} \varphi_0 \\ \theta_0 \end{pmatrix} = B \begin{pmatrix} \varphi_0 \\ \theta_0 \end{pmatrix},
\end{equation}
where we have introduced the matrix
\begin{equation}
 B = \begin{pmatrix}
  -1 & 1 \\ -1-h & 1
 \end{pmatrix}.
\end{equation}

The eigenvalues of $B$ are $\sigma = i \sqrt{h}$ and $\sigma^{\star} = - i \sqrt{h}$ (the star denotes the complex conjugate); the associated eigenvectors are $\Psi$ and $\Psi^{\star}$, where:
\begin{equation}
 \Psi = \begin{pmatrix} 1 \\ a \end{pmatrix},
\end{equation}
and where we have defined $a = 1 + i \sqrt{h}$. The solutions are thus:
\begin{equation} \label{eq:WW_weakly_nonlinear_solutions}
\begin{pmatrix}
 \varphi_0 \\ \theta_0
\end{pmatrix} = A(T) \Psi e^{i \sqrt{h} t} + A^{\star}(T) \Psi^{\star} e^{-i \sqrt{h} t}.
\end{equation}

\subsection{Order $1$}

At order $1$, the equation for $R$ is:

\begin{equation}
 \partial_t R_1 + R_1 = 1 - \frac{\Pi_c}{2} \left( \theta_0 - \varphi_0 \right)^2,
\end{equation}

where we have used that $\partial_T R_0 = 0$. We can compute:
\begin{align}
  \left( \theta_0 - \varphi_0 \right)^2 &= \left( i \sqrt{h} A e^{i \sqrt{h} t} + \text{c.c.} \right)^2 \nonumber\\
  &= 2h |A|^2 - \left( h A^2 e^{2i \sqrt{h}t} + \text{c.c.} \right),
\end{align}
where c.c. denotes the complex conjugate of the preceding term. 
We can write the solution for $R_1$ as:
\begin{equation}
 R_1 = 1 - h \Pi_c |A|^2 + \rho e^{2i\sqrt{h}t} + \text{c.c.},
\end{equation}
where $\rho$ is solution of $\left(1 + 2 i \sqrt{h}\right)\rho = h \Pi_c A^2 /2$. 
Finally, we find
\begin{equation}
 R_1 = 1 - h \Pi_c |A|^2 + \frac{h \Pi_c A^2}{2\left( 1 + 2 i \sqrt{h} \right)} e^{2i\sqrt{h}t} + \text{c.c.}.
\end{equation}

\begin{widetext}

\subsection{Order $3/2$}

To determine the equation for $\theta$ at order $3/2$, we need the expansion of $\Pi/R$:

\begin{equation}
 \frac{\Pi}{R} \simeq \frac{\Pi_c + \varepsilon}{\Pi_c + \varepsilon R_1} \simeq 1 + \varepsilon \frac{1 - R_1}{\Pi_c}.
\end{equation}

The equations for $\varphi$ and $\theta$ are thus:

\begin{subequations}
\begin{align}
\partial_t \varphi_1 + \varphi_1 - \theta_1 &= - \partial_T \varphi_0 + \frac{1 - R_1}{\Pi_c} \left( \theta_0 - \varphi_0 \right) - \frac{1}{6} \left( \theta_0 - \varphi_0 \right)^3, \\
\partial_t \theta_1 + (1+h)\varphi_1 - \theta_1 &= - \partial_T \theta_0 + R_1 (\theta_0 - \varphi_0) - \frac{\Pi_c}{6} \left( \theta_0 - \varphi_0 \right)^3 + \frac{h}{6} \theta_0^3.
\end{align}
\end{subequations}

We see that $\varphi_1$ and $\theta_1$ are driven by terms that are at their resonance frequencies $\pm \sqrt{h}$, which may cause a divergence. Assuming that there is a solution gives a condition on the right-hand side. To see this, we write these equations as

\begin{equation}
 \partial_t \begin{pmatrix} \varphi_1 \\ \theta_1 \end{pmatrix} - B \begin{pmatrix} \varphi_1 \\ \theta_1 \end{pmatrix} = L
\end{equation}

Now restricting ourselves to terms at the frequency $\omega = \sqrt{h}$, which we denote by $\left[ \cdot \right]_{\sqrt{h}}$, we get

\begin{equation}
 \left( i \sqrt{h} - B \right) \left[ \begin{pmatrix} \varphi_1 \\ \theta_1 \end{pmatrix} \right]_{\sqrt{h}} = \left[ L \right]_{\sqrt{h}}
\end{equation}

This is an equation of the form $(i \sqrt{h} - B)X = C$, so that the solution could be looked for as the combination $X = \lambda \Psi + \lambda_{\star} \Psi^{\star}$. Inserting this decomposition in the equation, and using that $B\Psi = i\sqrt{h}\Psi$ and $B \Psi^{\star} = -i\sqrt{h} \Psi^{\star}$, we obtain $2i\sqrt{h} \lambda_{\star} \Psi^{\star} = C$, meaning that $C$ should be colinear to $\Psi^{\star}$. This condition can be written with the determinant: $C_{\theta} - a^{\star} C_{\varphi} = 0$, where $C_{\varphi}$ and $C_{\theta}$ are the components of $C = \left[ L \right]_{\sqrt{h}}$, which we now determine explicitly.

For the derivatives with respect to $T$, we obtain:
\begin{subequations}
\begin{align}
\left[ \partial_T \varphi_0 \right]_{\sqrt{h}} &= \dot{A}, \\
\left[ \partial_T \theta_0 \right]_{\sqrt{h}} &= a\dot{A}.
\end{align}
\end{subequations}
We have
\begin{equation}
 \left[ \theta_0 - \varphi_0 \right]_{\sqrt{h}} = i \sqrt{h} A,
\end{equation}
and
\begin{subequations}
\begin{align}
 \left[ R_1(\theta_0 - \varphi_0) \right]_{\sqrt{h}} &= \left[ R_1 \right]_0 \left[ \theta_0 - \varphi_0 \right]_{\sqrt{h}} + \left[ R_1 \right]_{2\sqrt{h}} \left[ \theta_0 - \varphi_0 \right]_{-\sqrt{h}} \\
 &= \left( 1 - h \Pi_c |A|^2 \right) i \sqrt{h} A + \frac{h \Pi_c A^2}{2(1 + 2i\sqrt{h})} \left( -i\sqrt{h} A^{\star} \right) \\
 &= i\sqrt{h}A - \frac{ih\sqrt{h}\Pi_c}{2} \left( 2 + \frac{1}{1 + 2i\sqrt{h}} \right) |A|^2 A.
\end{align}
\end{subequations}

The cubic terms are given by
\begin{equation}
 \left[ (\theta_0 - \varphi_0)^3 \right]_{\sqrt{h}} = 3 \left[ \theta_0 - \varphi_0 \right]^2_{\sqrt{h}} \left[ \theta_0 - \varphi_0 \right]_{-\sqrt{h}} = 3 \left( i\sqrt{h}A \right)^2 \left( -i\sqrt{h}A^{\star} \right) = 3 i h \sqrt{h} |A|^2 A,
\end{equation}
and
\begin{equation}
 \left[ \theta_0^3 \right]_{\sqrt{h}} = 3 \left[ \theta_0 \right]^2_{\sqrt{h}} \left[ \theta_0 \right]_{-\sqrt{h}} = 3 \left( a A \right)^2 (a^{\star}A^{\star}) = 3a\Pi_c |A|^2 A.
\end{equation}
We have used that $|a|^2 = 1 + h = \Pi_c$.

We can compile these terms:
\begin{subequations}
 \begin{align}
C_{\varphi} &= - \dot{A} + \frac{ih\sqrt{h}}{2} \left( 2 + \frac{1}{1 + 2i\sqrt{h}} \right) |A|^2 A - \frac{ih\sqrt{h}}{2} |A|^2 A \\
&= - \dot{A} + i h \sqrt{h} \frac{ 1 + i \sqrt{h} }{ 1 + 2i\sqrt{h} } |A|^2 A.
 \end{align}
\end{subequations}
For $C_{\theta}$, we get
\begin{subequations}
 \begin{align}
C_{\theta} &= -a\dot{A} + i\sqrt{h}A - \frac{ih\sqrt{h}\Pi_c}{2} \left( 2 + \frac{1}{1 + 2i\sqrt{h}} \right) |A|^2 A - \frac{ih\sqrt{h}\Pi_c}{2} |A|^2 A + \frac{a h \Pi_c}{2} |A|^2 A \\
&= -a\dot{A} + i\sqrt{h} A + \frac{h\Pi_c(4h - i\sqrt{h} + 1)}{2(1 + 2i\sqrt{h})} |A|^2 A.
 \end{align}
\end{subequations}

The equation $C_{\theta} - a^{\star} C_{\varphi} = 0$ thus reads, after simplification:
\begin{equation}
 \dot{A} = \frac{A}{2} - Z |A|^2 A,
\end{equation}
where
\begin{equation}
 Z = \frac{ (h+1)\left( 8h^2 + 5h - 2ih\sqrt{h} + i \sqrt{h} \right) }{ 4(4h+1) }.
\end{equation}

The real part of $Z$, $Z_r$, sets the amplitude:
\begin{equation} \label{eq:final_result_WW_weakly_nonlinear}
 |A| = \frac{1}{\sqrt{2 Z_r}} = \sqrt{ \frac{ 2(4h+1) }{ h(h+1)(8h + 5) } }
\end{equation}
The imaginary part of $Z$ introduces a correction to the oscillating frequencies $\pm \sqrt{h}$. Moreover, we can obtain the asymptotic behaviors of the amplitude:
\begin{subequations}
\begin{align}
  |A| &\underset{h \to 0}{\sim} \sqrt{ \frac{2}{5h} },\\
 |A| &\underset{h \to \infty}{\sim} \frac{1}{h}.
\end{align}  
\end{subequations}
These behaviors are expected: as $h \rightarrow \infty$, the motion is more and more constrained by the external field and the amplitude decays. 
On the contrary, as $h \rightarrow 0$ the system is closer and closer to the exceptional point and the amplitude diverges.

\subsection{Numerical simulations}

In the numerical simulations, we observe that the angle $\theta(t)$ follows $\theta(t) = A_{\theta} \cos( \omega t )$ close to the instability threshold. From the above analysis, it follows that $A_{\theta} = 2 \sqrt{1 + h} \sqrt{\varepsilon} A$, where $A$ is the amplitude computed above (Eq. (\ref{eq:final_result_WW_weakly_nonlinear})). The factor $2$ comes from the fact that we add the complex conjugate and take the real part. The factor $\sqrt{1 + h}$ comes from the $\Psi$ factor in Eq. (\ref{eq:WW_weakly_nonlinear_solutions}), whose $\theta$ component is $1 + i \sqrt{h}$. Finally, the factor $\sqrt{\varepsilon}$ comes from the expansion of $\theta$ (Eq. (\ref{eq:WW_weakly_nonlinear_expansion_theta})). 
Hence, the theoretical prediction is
\begin{equation}\label{eq:single_wnl_At}
 A_{\theta} = \sqrt{ \frac{ 8 \varepsilon (4h + 1) }{ h(8h + 5) } }.
\end{equation}
This prediction is compared to the results of numerical simulations in Fig.~\ref{fig:single_wnl}; an excellent agreement is obtained.

\begin{figure}
\centering
\includegraphics[scale=.9]{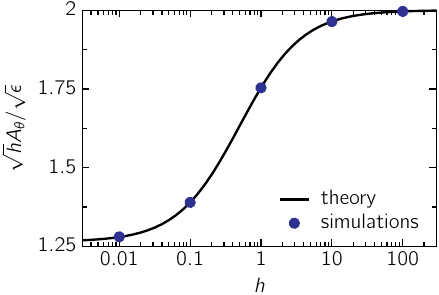}
\caption{Amplitude $A_\theta$ of the oscillations along $\theta$ close to the FP-WW transition; solid line: theoretical prediction as given by Eq.~(\ref{eq:single_wnl_At}); blue markers: numerical simulations.}
\label{fig:single_wnl}
\end{figure}

\subsection{Expansion close to the exceptional point}

To compare the expansion in the weakly nonlinear WW regime with the expansion close to the exceptional point, we specify our results to the case $h\to 0$.
We find in this limit that
\begin{equation}
    \varphi(t)
    \simeq \theta(t)\simeq \sqrt{\frac{8}{5} \frac{\Pi-1-h}{h}} \cos\left(\sqrt{h} t\right) .
\end{equation}
Noting that we have taken $\omega_0^2=1$ here, this result matches the one obtained from an expansion close to the exceptional point in the small amplitude limit, Eq.~(\ref{eq:exp_exceptional_comp}).

\section{Single particle in the degenerate case: the CO regime in the small field limit}
\label{app:linear_response_CO}

Here, we find the linear response of the CO regime to a small external field. Starting from Eqs. (\ref{eq:1part_degenerated}), we linearize the dynamics around the zero-field CO regime ($R_0 = \sqrt{\Pi}/\omega_{0}$, $\cos \gamma_0 = \omega_0/\sqrt{\Pi}$, $\dot{\varphi} = \Omega_0 = \omega_0\sqrt{\Pi - \omega_{0}^{2}}$), and introduce the small quantities $R(t) = R_0 + \delta R(t)$, $\gamma(t) = \gamma_0 + \delta \gamma$, $\varphi = \Omega_0 t + \delta \varphi$, $h = \delta h$. We find:
\begin{equation} \label{eq:LR_CO}
\frac{d}{dt}\begin{pmatrix}
 \delta R \\
 \delta \varphi \\
 \delta \gamma
\end{pmatrix} =
\begin{pmatrix}
 - \omega_{0}^{2} & 0 & - \Pi \sin \gamma_0 \\
 - \omega_{0}^{2} \sin \gamma_0 & 0 & \omega_{0}^{2} \\
 2 \omega_{0}^{2} \sin \gamma_0 & 0 & 0 \\
\end{pmatrix}
\begin{pmatrix}
\delta R \\
\delta \varphi \\
\delta \gamma
\end{pmatrix}
+ \delta h \sin(\gamma_0 + \Omega_0 t)
\begin{pmatrix}
0 \\
0 \\
1
\end{pmatrix},
\end{equation}
where the matrix on the r.h.s. not only allows to access the stability of the CO regime, but also the linear response of this regime to a small external field. We find that assessing the stability reduces to the following eigenvalue problem:
\begin{equation} \label{eq:PCO_eigenvalues}
 \lambda \left[ \lambda^2 + \lambda \omega_{0}^{2} + 2\omega_{0}^{2} \left( \Pi - \omega_{0}^{2} \right) \right] = 0,
\end{equation}
which only has one zero solution in the $\delta \varphi$ direction (the phase of the oscillation is marginally stable). We denote $\Delta = \omega_{0}^{4} - 8\omega_{0}^{2}\left( \Pi - \omega_{0}^{2} \right)$ the determinant of Eq. (\ref{eq:PCO_eigenvalues}) and $\delta \Pi = \Pi - \omega_{0}^{2}$ the distance to threshold. For $\delta \Pi / \omega_{0}^{2} < 1/8$, the two remaining eigenvalues are reals and negatives: $\lambda = ( - \omega_{0}^{2} \pm \sqrt{\Delta})/2$; and for $\delta \Pi / \omega_{0}^{2} > 1/8$, they are complex conjugates with negative real parts: $\lambda = ( - \omega_{0}^{2} \pm i \sqrt{-\Delta})/2$. Thus, the zero-field CO regime is stable in its whole range of existence. Introducing the complex amplitudes $A_R$, $A_{\varphi}$ and $A_{\gamma}$, we look for solutions of the form ($\delta R$, $\delta \varphi$, $\delta \gamma$) = ($A_R$, $A_{\varphi}$, $A_{\gamma}$)$e^{i\Omega_0 t}$. We find the following condition:
\begin{equation} \label{eq:LR_CO_ansatz}
\begin{pmatrix}
 - \omega_{0}^{2} -i\Omega_0 & 0 & - \Pi \sin \gamma_0 \\
 - \omega_{0}^{2} \sin \gamma_0 & -i\Omega_0 & \omega_{0}^{2} \\
 2 \omega_{0}^{2} \sin \gamma_0 & 0 & -i\Omega_0 \\
\end{pmatrix} \begin{pmatrix}
 A_R \\
 A_{\varphi} \\
 A_{\gamma}
\end{pmatrix} =
A_h
\begin{pmatrix}
0 \\
0 \\
1
\end{pmatrix},
\end{equation}
where $A_h = \delta h e^{i\Psi}$, and $\Psi$ is an irrelevant phase shift. The matrix on the left-hand side of Eq. (\ref{eq:LR_CO_ansatz}) is invertible for $\Pi > \omega_{0}^{2}$. At lowest order in $\delta \Pi = \Pi - \omega_{0}^{2}$, the complex amplitudes write:
\begin{equation} \label{eq:LR_CO_ansatz_sol}
\begin{pmatrix}
 | A_{R} | \\
 | A_{\varphi} | \\
 | A_{\gamma} |
\end{pmatrix} \simeq
| A_h | \begin{pmatrix}
1/\omega_{0}^{2} \\
1/ \delta \Pi \\
1/\omega_{0}\sqrt{\delta \Pi}
\end{pmatrix}.
\end{equation}
The complex amplitudes for the modulations along $\varphi$ and $\gamma$ diverge as one gets closer to the exceptional point, which induces a change of regime.

\end{widetext}

\end{document}